\documentclass[prd,twocolumn,nofootinbib,superscriptaddress]{revtex4-2}
\usepackage{bm}
\usepackage{amsmath}
\usepackage{amssymb}
\usepackage{graphicx}
\usepackage{subfigure}
\usepackage{hyperref}
\usepackage{color}
\usepackage{comment}
\usepackage{ulem}
\usepackage{CJK}
\hypersetup{
	colorlinks=true,
	linkcolor=red,
	citecolor=blue,
}
\allowdisplaybreaks[2]

\begin{document}

\begin{CJK*}{UTF8}{gbsn}
\title{Accuracy of parameter estimations with space-borne gravitational wave observatory} 	
\author{Chao Zhang (张超)}
\email{chao\_zhang@hust.edu.cn}
\affiliation{School of Physics, Huazhong University of Science and Technology,
Wuhan, Hubei 430074, China}
\author{Yungui Gong (龚云贵)}
\email{Corresponding author. yggong@hust.edu.cn}
\affiliation{School of Physics, Huazhong University of Science and Technology, Wuhan, Hubei 430074, China}
\author{Bin Wang (王斌)}
\email{wang\_b@sjtu.edu.cn}
\affiliation{Center for Gravitation and Cosmology, Yangzhou University, Yangzhou 225009, China}
\author{Chunyu Zhang (张春雨)}
\email{chunyuzhang@hust.edu.cn}
\affiliation{School of Physics, Huazhong University of Science and Technology,
Wuhan, Hubei 430074, China}

\begin{abstract}
Employing the Fisher information matrix analysis,
we estimate parameter errors of TianQin and LISA for monochromatic gravitational waves.
With the long-wavelength approximation we derive analytical formulas for the parameter estimation errors.
We separately analyze the effects of the amplitude modulation due to the changing orientation of the detector plane and the Doppler modulation due to the translational motion of the center of the detector around the Sun.
We disclose that in the low frequency regime there exist
different patterns in angular resolutions and estimation errors of sources' parameters between LISA and TianQin,
the angular resolution  falls off as $S_n(f)/f^2$ for TianQin but $S_n(f)$ for LISA,
and the estimation errors of the other parameters fall off as $\sqrt{S_n(f)}/f$ for TianQin but $\sqrt{S_n(f)}$ for LISA.
In the medium frequency regime we observe the same pattern where the angular resolution falls off as $S_n(f)/f^2$ and the estimation errors of the other parameters  fall off as $\sqrt{S_n(f)}$ for both TianQin and LISA.
In the high frequency regime, the long-wavelength approximation fails, we numerically calculate the parameter  estimation errors for LISA and TianQin and find that the parameter estimation errors measured by TianQin are smaller than those by LISA.

\end{abstract}

\preprint{2012.01043}

\maketitle
\end{CJK*}

\section{Introduction}

Tens of gravitationanl wave (GW) detections in the frequency range  tens to hundreds of hertz were reported
by the Laser Interferometer Gravitational-Wave Observatory (LIGO) Scientific Collaboration and the Virgo Collaboration \cite{TheLIGOScientific:2016agk,Abbott:2016blz,Abbott:2016nmj, Abbott:2017vtc,Abbott:2017oio,TheLIGOScientific:2017qsa, Abbott:2017gyy,LIGOScientific:2018mvr,Abbott:2020uma,LIGOScientific:2020stg,Abbott:2020khf,Abbott:2020tfl,Abbott:2020niy}.
It is expected that the proposed space-based observatories such as
Laser Interferometer Space Antenna (LISA) \cite{Danzmann:1997hm,Audley:2017drz}, Taiji \cite{Hu:2017mde} and TianQin  \cite{Luo:2015ght}
will detect GWs in the frequency regimes $10^{-4}-10^{-1}$ Hz
which are undetectable by the ground-based GW observatories due to the seismic noise and gravity gradient noise.

There are two different configurations and designs for the space-based GW observatory.
One configuration has the geocentric orbit  with three spacecrafts orbiting the Earth and further rotating around the Sun together with the earth.
TianQin takes such design  with
the normal vector of the detector plane pointing
to the source RX J0806.3+1527 \cite{Luo:2015ght}. LISA \cite{Danzmann:1997hm,Audley:2017drz} and Taiji \cite{Hu:2017mde} adopt the design with the three spacecrafts moving in  the heliocentric orbit ahead or behind the Earth by about $20^\circ$.
LISA and Taiji constellations have an inclination angle of $60^\circ$ with respect to the ecliptic plane.
The inclination angle ensures the three spacecrafts to keep the geometry of an equilateral triangle throughout the mission.
The normal vector of the detector plane rotates around the normal vector of the ecliptic plane and
forms a cone with a half opening angle of $60^\circ$ in one year.

Each measured GW signal shaped by a set of parameters carries information about the source:
its location in the sky, the inclination angle between the binary's orbital angular momentum and the line of sight,
the polarization angle, the amplitude, the initial phase and the frequency which can be accurately determined by the orbital period.
The physical parameters of the binary source extracted from the GW signal are strongly correlated.
The ability to locate a source in the sky is important  in
GW observations and parameter estimations, therefore
localizing the sky position of the GW source is a key scientific goal for GW observations.
The accuracy of the source localization is essential
for the follow-up observations of counterparts and the statistical
identification of the host galaxy when no counterpart is present.
With the information about counterparts or the host galaxy,
the redshift $z$ of the source  can be determined
and unprecedented information about its environment may be uncovered.
Combining the redshift $z$ with the luminosity distance $d_L$ to the
source determined from the GW waveform, we can probe
the thermal history of the Universe and measure fundamental cosmological parameters.
In particular, the independent measurements of $d_L$ and $z$ from GWs can be used
as standard sirens \cite{Schutz:1986gp,Holz:2005df} to understand the problem of Hubble tension \cite{Riess:2019cxk}.

Except for different frequency bands between space-based and ground-based detectors,
another important difference is that space-based GW observatories
can observe binary sources for months to years before the final coalescence.
Thus for space-based observations, the periodic Doppler shift imposed on the signal
by the translational motion of the detector around the Sun
results in amplitude and phase modulations of the detector output
which encode information about both the detector position and the angular position of the source, hence the periodic Doppler shift
provides us a method of identifying the angular position
of the source in the sky with one detector.
The extraction of parameters from merging compact binaries for LIGO and Virgo network was first discussed in \cite{Cutler:1994ys}.
The accuracy of the angular resolution of LISA for a monochromatic GW was
first investigated in \cite{Peterseim:1996cw,Peterseim:1997ic}  with the simplified detector response and the assumption that all other parameters are known a prior.
Cutler then estimated the angular resolution of LISA and all other parameters including the frequency simultaneously for galactic and extragalactic sources by using a more realistic detector response which accounts for the rotation of the LISA constellation in the long-wavelength approximation \cite{Cutler:1997ta,Cutler:1998muh}.
More accurate waveforms were also included  in the context of coalescing massive black hole (MBH) binaries and captures of small compact objects by MBHs \cite{Moore:1999zw,Barack:2003fp,Porter:2008kn,Blaut:2011zz}.
For equal mass black hole (BH) binary system with the total mass $10^5 M_\odot$
at the redshift $z=1$,
the LISA-Taiji network can improve the accuracy of the sky localization
by two orders of magnitude than each individual detector for one year's joint observation,
and an optimal configuration angle of $40^\circ$ was suggested for the LISA-Taiji network \cite{Ruan:2019tje,Ruan:2020smc}.
By simulating coalescing GWs from MBH binary systems with masses $M_1=10^7 M_\odot$,
$10^6 M_\odot$ and $10^5 M_\odot$ respectively for the primary BH
and the mass ratio $q=1/3$ at the redshift $z=2$,
and 30 days observation until the merger,
it was found that the LISA-Taiji network could improve
the angular resolution for various time-delay interferometry channels
by more than 10 times than each individual LISA or Taiji detector \cite{Wang:2020vkg}.
The improvement from the LISA-Taiji network is relatively moderate
for monochromatic GWs at 3 mHz and 10 mHz with one year observation \cite{Wang:2020vkg}.
The precisions of the parameter estimation and the sky localization of
equal mass MBH binary systems with masses in the range $10^5-10^7M_\odot$
for TianQin were discussed in \cite{Feng:2019wgq}.
The LISA-TianQin network may improve the sky localization of
Galactic double white dwarf binaries up to 3 orders of magnitude \cite{Huang:2020rjf} comparing with single TianQin detector.
The effects on the sky localization due to the different configuration between LISA/Taiji and TianQin and their sky maps of the sky localization for monochromatic GWs
were analyzed in \cite{Zhang:2020hyx}.
For more discussion on the accuracy of sky localization,
please see Refs. \cite{Vallisneri:2007ev,Wen:2010cr,Aasi:2013wya,Grover:2013sha,Berry:2014jja,Singer:2015ema,Becsy:2016ofp,Zhao:2017cbb,Mills:2017urp,Fairhurst:2017mvj,Fujii:2019hdi}.

In this paper, we consider GW signals from two compact objects such as white dwarfs, neutron stars or black holes.
During the operation time of the space-borne GW observatory,
there is nearly no frequency evolution in the early inspiral of these systems.
The GWs emitted by these sources can be treated as monochromatic.
The detection of these sources is anticipated to provide us with new insights
into the formation and the evolution of relativistic objects
and the physics in the early universe.
The aim of the present paper is to investigate the accuracy of parameter estimations for TianQin and LISA.
The analysis can be easily applied to Taiji because of its similarity with LISA.
Estimation errors of sky positions of sources and the other parameters are mainly dependent on the amplitude modulation due to the changing orientation of the detector plane and the Doppler effect due to the translational motion of the center of the detector around the Sun.
The amplitude of the detector response is modulated by the annual rotation of the LISA array which improves the measurement of the sky position.
For TianQin, there exists no such modulation and its sky localization is expected to be different from LISA.
Apart from the amplitude modulation,
the phase of the detector response modulated by the Doppler effect for TianQin and LISA also improves the accuracy of parameter estimation.
Therefore, the correlation between the amplitude modulation and Doppler modulation makes the analysis of sky localization quite different between TianQin and LISA.

We present detailed analysis of the estimation of the parameters with TianQin and LISA for monochromatic GWs.
To understand the result, the exact detector signal and its long-wavelength approximation are considered.
In the long-wavelength regime we give compact formulas for the errors as functions of the frequency.
With these analytical formulas,
it is easy to understand the frequency dependence of the parameter estimation errors in the low and medium frequency regimes.
In the high frequency domain we numerically calculate errors as functions of the frequency for the single Michelson observable.
The organization of the paper is as follows.
In Sec. \ref{sec2}, we review the Fisher Information Matrix (FIM) method of signal analysis.
In Sec. \ref{sec3}, we apply the FIM method to TianQin and LISA.
We derive the analytical formulas of the parameter estimation errors in the long-wavelength domain and we put
the detailed analyses in the Appendices \ref{AppendixTianQin}, \ref{AppendixLISA} and \ref{InverseFIM};
in high frequency regime, we give the numerical results of the parameter estimation errors.
For comparison, we consider the parameter estimation errors for four parameters in Appendix  \ref{TQ4paras} and the Bayesian analysis for LISA in Appendix  \ref{bayes1}.
We present our conclusion and discussion in the last section.

\section{The Fisher Information Matrix Method}
\label{sec2}

It is convenient to describe GWs and the motion of detectors in the heliocentric right-handed orthogonal reference frame with the constant basis vectors $\{\hat{e}_x,\hat{e}_y,\hat{e}_z\}$ \cite{Rubbo:2003ap}.
For GWs propagating in the direction $\hat{w}$,
we introduce a set of unit vectors $\{\hat{w}, \hat{\theta}, \hat{\phi} \}$ which are perpendicular to each other as
\begin{equation}\label{polarization}
\begin{aligned}
\hat{\theta}&=\cos(\theta)\cos(\phi)\hat{e}_x+ \cos(\theta)\sin(\phi) \hat{e}_y-\sin(\theta)\hat{e}_z,\\
\hat{\phi}&=-\sin(\phi)\hat{e}_x+\cos(\phi)\hat{e}_y,\\
\hat{w} &=-\sin(\theta)\cos(\phi)\hat{e}_x -\sin(\theta)\sin(\phi)\hat{e}_y -\cos(\theta)\hat{e}_z,
 \end{aligned}
\end{equation}
where the angles $\theta$ and $\phi$ are the angular coordinates of the source.
To describe GW signals, the polarization angle $\psi$ is introduced to form polarization tensors
\begin{equation}
\begin{split}
e^{+}_{ij}=\hat{p}_i\hat{p}_j-\hat{q}_i\hat{q}_j,& \quad e^{\times}_{ij}=\hat{p}_i\hat{q}_j+\hat{q}_i\hat{p}_j,\\
\hat{p}=\cos\psi \hat{\theta}+\sin \psi \hat{\phi},& \quad \hat{q}=-\sin\psi \hat{\theta}+\cos\psi \hat{\phi}.
\end{split}
\end{equation}
where $\hat{p}$ and $\hat{q}$ represent the directions of the two polarization axes of the gravitational radiation.
For this particular choice of $\{ \hat{w}, \hat{p},\hat{q}\}$,  GWs have the form
\begin{equation}
h_{ij}(t)=\sum_{A}e^A_{ij}h_A(t), \\
\end{equation}
where $A=+,\times$ stands for the plus and cross polarization states,
\begin{equation}
\begin{split}
h_+=&\mathcal{A}\left[1+\cos^2(\iota)\right]\exp (2\pi i f t + i\phi_0),\\
h_\times=&2i\mathcal{A}\cos(\iota)\exp (2\pi i f t + i\phi_0),
\end{split}
\end{equation}
$\mathcal{A}$ and $f$ are the amplitude and the frequency of GWs respectively,
$\iota$ is the inclination angle
and $\phi_0$ is the initial phase.

For a monochromatic GW with the frequency $f$ propagating along the direction $\hat{w}$,
the output of an equal arm space-based interferometric detector  with a single round trip of light travel  is
\begin{equation}
\label{gwst}
H(t)=\sum_A F^A h_A(t)e^{i\phi_D(t)},
\end{equation}
where $F^A=\sum_{i,j} D^{ij} e^A_{ij}$, the Doppler phase $\phi_D(t)$ is
\begin{equation}
\label{doppler}
   \phi_D(t)=\frac{2\pi  f R}{c}\sin\theta\cos\left(\frac{2\pi t}{T} -\phi-\phi_{\alpha}\right),
\end{equation}
$\phi_{\alpha}$ is the ecliptic longitude of the detector $\alpha$ at $t=0$,
$c$ is the speed of light,
the rotational period $T$ is 1 year and the radius $R$ of the orbit is 1 AU.
The detector tensor $D^{ij}$  is
\begin{equation}
D^{ij}=\frac{1}{2}[\hat{u}^i \hat{u}^j T(f,\hat{u}\cdot\hat{\omega})-\hat{v}^i \hat{v}^j  T(f,\hat{v}\cdot\hat{\omega})],
\end{equation}
where $\hat{u}$ and $\hat{v}$ are the unit vectors along the arms of the detector and $T(f,\hat{u}\cdot\hat{\omega})$ is \cite{Cornish:2001qi,Estabrook:1975}
\begin{equation}
\label{transferfunction}
\begin{split}
T(f,\hat{u}\cdot\hat{w})=&\frac{1}{2}
\left\{\text{sinc}[\frac{f(1-\hat{u}\cdot\hat{\omega})}{2f^*}]\exp[\frac{f(3+\hat{u}\cdot\hat{\omega})}{2if^*}] \right.\\
+&\left.\text{sinc}[\frac{f(1+\hat{u}\cdot\hat{\omega})}{2f^*}]\exp[\frac{f(1+\hat{u}\cdot\hat{\omega})}{2if^*}]\right\},
\end{split}
\end{equation}
$\text{sinc}(x)=\sin x/x$, 
$f^*=c/(2\pi L)$ is the transfer frequency of the detector 
and $L$ is the arm length of the detector.

The signal $H(t)$ depends on the source parameters which are to be estimated.
To estimate the accuracy of the parameters by a single detector,
we introduce the FIM in the frequency domain
\begin{equation}
\label{gijproduct}
\begin{split}
\Gamma_{ij}=& \left(\frac{\partial H}{\partial \xi_i}\left|\frac{\partial H}{\partial \xi_j}\right.\right)\\
=&4\Re \int_{0}^{\infty}\frac{\partial_i H(f)\partial_j H^*(f)}{S_{n}(f)}df,
\end{split}
\end{equation}
where $\Re$ means the real part, $\partial_i H=\partial H/\partial  \xi_i$
and $\xi_i$ is the $i$-th parameter.
For monochromatic GW sources there is almost no frequency evolution, the FIM can be simplified in the time domain as
\begin{equation}
\label{gamma}
\Gamma_{ij}
=\frac{2}{S_{n}(f)}\Re\int_{0}^{T_{obs}}\partial_i H(t)\partial_j H^*(t)dt,
\end{equation}
where  $T_{obs}=1$ yr is the observational time.
We focus mainly on the following  parameters of the monochromatic GW signal considered in Eq. \eqref{gamma},
\begin{equation}
\label{parameters}
    {\bm \xi}=\{ \theta, \phi,\ln{\mathcal{A}},\iota,\psi,\phi_0\}.
\end{equation}
For space-based interferometers, the noise power spectral density $S_n(f)$ is \cite{Audley:2017drz}
\begin{equation}\label{Sn}
S_n(f)=\frac{S_x}{L^2}+\frac{4S_a}{(2\pi f)^4L^2}\left(1+\frac{10^{-4}\text{Hz}}{f}\right).
\end{equation}
For LISA, the acceleration noise is $\sqrt{S_a}=3\times 10^{-15}~ \text{m s}^{-2}/\text{Hz}^{1/2}$,
the displacement noise is $\sqrt{S_x}=15\ \text{pm/Hz}^{1/2}$, the arm length is $L_s=2.5 \times 10^6$ km \cite{Audley:2017drz},
and its transfer frequency is $f^*_s=0.02$ Hz.
For TianQin, the acceleration noise is $\sqrt{S_a}=10^{-15}\
\text{m s}^{-2}/\text{Hz}^{1/2}$,
the displacement noise is $\sqrt{S_x}=1\ \text{pm/Hz}^{1/2}$,
the arm length is $L_t=1.7\times 10^5$ km \cite{Luo:2015ght},
and its transfer frequency is $f^*_t=0.28$ Hz.

The FIM can be split into three parts.
$\Gamma^{am}_{ij}$ denotes the total amplitude modulation  and  $\Gamma^{dm}_{ij}$ denotes the Doppler phase modulation and  $\Gamma^{ad}_{ij}$ denotes the interaction of the total amplitude and Doppler phase  modulation
\begin{equation}\label{gammaij}
\Gamma_{ij}=\Gamma^{am}_{ij}+\Gamma^{dm}_{ij}+\Gamma^{ad}_{ij},
\end{equation}
where
\begin{equation*}
\begin{split}
\Gamma^{am}_{ij}=& \left(\frac{\partial \left(\sum_A F^A h_A\right)}{\partial \xi_i}\left|\frac{\partial \left(\sum_A F^A h_A\right)}{\partial \xi_j}\right.\right),\\
\end{split}
\end{equation*}
\begin{equation*}
\begin{split}
\Gamma^{dm}_{ij}=&\left(\sum_A F^A h_A\frac{\partial \left(e^{i\phi_D(t)}\right)}{\partial \xi_i}\left|\sum_A F^A h_A\frac{\partial \left(e^{i\phi_D(t)}\right)}{\partial \xi_j}\right.\right),\\
\end{split}
\end{equation*}
\begin{equation*}
\begin{split}
\Gamma^{ad}_{ij}=&  \left(\frac{\partial \left(\sum_A F^A h_A\right)}{\partial \xi_i}e^{i\phi_D(t)}\left|\sum_A F^A h_A\frac{\partial \left(e^{i\phi_D(t)}\right)}{\partial \xi_j}\right.\right)\\
+&  \left(\sum_A F^A h_A\frac{\partial \left(e^{i\phi_D(t)}\right)}{\partial \xi_i}\left|\frac{\partial \left(\sum_A F^A h_A\right)}{\partial \xi_j}e^{i\phi_D(t)}                                                    \right.\right).
\end{split}
\end{equation*}
The covariance matrix of the parameter errors is
\begin{equation}
\sigma_{ij}=\left\langle\Delta\xi^i\Delta\xi^j\right\rangle\approx (\Gamma^{-1})_{ij}.
\end{equation}
The root mean square errors of the parameters are given by
\begin{equation}
    \sigma_{i}=\sqrt{\sigma_{ii}}=\sqrt{(\Gamma^{-1})_{ii}}\,.
\end{equation}
The angular uncertainty of the sky localization is evaluated as \cite{Cutler:1997ta}
\begin{equation}
\Delta \Omega\equiv2\pi\sin\theta
\sqrt{\sigma_{\theta\theta}\sigma_{\phi\phi}-\sigma^2_{\theta\phi}}\,.
\end{equation}

\section{Parameter Estimation Errors}
\label{sec3}
In this section, we analyze the errors of the six parameters discussed in the previous section.
If the parameters $\psi$ and $\phi_0$ are known,
then the six parameters are reduced to four parameters, we discuss the results of the four parameters in Appendix \ref{TQ4paras}.

\subsection{The long-wavelength approximation}

In this subsection we derive analytical formulas for the parameter estimation errors in the long-wavelength (LW) approximation for LISA and TianQin.
We also use numerical method to confirm the analytical behaviours.
Note that in the LW approximation, $f\ll f^*$,
and $T(f, \hat{u}\cdot\hat{w})\rightarrow$ 1 in Eq. (\ref{transferfunction}).

TianQin is an equilateral triangle constellation with sides of $1.73\times10^5$ km designed to orbit the Earth with the period of $3.65$ days and further rotate around the Sun together with the Earth \cite{Luo:2015ght}.
In the heliocentric coordinate system  the normal vector of the detector plane points to the direction of RX J0806.3+1527 with the latitude $\beta=94.7^\circ$ and the longitude $\alpha=120.5^\circ$. The orbits of the unit vectors of detector arms (two arms only) for TianQin are \cite{Hu:2018yqb}
\begin{equation*}
\label{detectorTQ}
\begin{aligned}
\hat{u}_x&=\cos(\omega_s t)\cos(\alpha)\cos(\beta)-\sin(\omega_st)\sin(\alpha),\\
\hat{u}_y&=   \cos(\alpha)\sin(\omega_st)+\cos(\omega_st)\cos(\beta)\sin(\alpha),\\
\hat{u}_z&=-\cos(\omega_st)\sin(\beta),\\
\hat{v}_x&=\cos(\omega_st+\frac{\pi}{3})\cos(\alpha)\cos(\beta)
-\sin(\omega_st+\frac{\pi}{3})\sin(\alpha),\\
\hat{v}_y&=   \cos(\alpha)\sin(\omega_st+\frac{\pi}{3})
+\cos(\omega_st+\frac{\pi}{3})\cos(\beta)\sin(\alpha),\\
\hat{v}_z&=-\cos(\omega_st+\frac{\pi}{3})\sin(\beta),\\
\end{aligned}
\end{equation*}
where the rotation frequency $\omega_s=2\pi/(3.65$ days).
The FIM in the LW approximation can be written as a sum in a compact form (the matrix elements are presented in Appendix \ref{AppendixTianQin})
\begin{equation}
\label{gammaijTQ}
 \Gamma^{LW}=\Gamma^{am}+\Gamma^{dm}+\Gamma^{ad},
\end{equation}
where
\begin{equation}
\label{gammaijTQ2}
\begin{split}
 \Gamma^{am}&=\frac{\mathcal{A}^2T_{obs}  }{S_n(f)}M^{am},\quad
  \Gamma^{dm}=\frac{\mathcal{A}^2 T_{obs} }{S_n(f)}M^{dm},\\
   \Gamma^{ad}&=\frac{\mathcal{A}^2T_{obs} }{S_n(f)}M^{ad},
\end{split}
\end{equation}
$M^{am}$ is singular and independent of the frequency,
\begin{equation}
 M^{dm}=\left(\frac{2\pi f R}{c} \right)^2\begin{pmatrix}
\textbf{D}_{2\times2}, \textbf{0}_{2\times4}\\
\textbf{0}_{4\times2},\textbf{0}_{4\times4}
\end{pmatrix},\qquad  M^{ad}=0.
\end{equation}
Note that only the submatrix $\bf{D}$ is nonzero,
the matrix $M^{dm}$ for the Doppler phase modulation depends on the frequency as $f^2$
and the matrix $M^{am}$ is independent of the frequency.

LISA mission was proposed as an equilateral triangle constellation with sides of $2.5\times10^6$ km \cite{Danzmann:1997hm,Audley:2017drz}.
The constellation has an inclination angle of $60^\circ$ with respect to the ecliptic plane and trails the Earth by about $20^\circ$.
In the heliocentric coordinate system, the orbits of the unit vectors of detector arms (two arms only) for LISA are \cite{Cutler:1998muh}
\begin{equation*}\label{detectorLISA}
\begin{aligned}
\hat{u}_x&=-\sin(\omega_st ) \cos(\omega_st)  + \cos(\omega_st) \sin(\omega_st)/2,             \\
\hat{u}_y&=     \cos(\omega_st)\cos(\omega_st)+\sin(\omega_st)  \sin(\omega_st)/2,        \\
\hat{u}_z&=\sin(\pi/3) \sin(\omega_st),         \\
\hat{v}_x&=-\sin(\omega_st ) \cos(\omega_st-\alpha)+ \cos(\omega_st) \sin(\omega_st-\alpha)/2,             \\
\hat{v}_y&=     \cos(\omega_st)\cos(\omega_st-\alpha)+ \sin(\omega_st)  \sin(\omega_st-\alpha)/2,        \\
\hat{v}_z&=\sin(\pi/3) \sin(\omega_st-\alpha),         \\
\end{aligned}
\end{equation*}
where the rotation frequency $\omega_s=2\pi/(365 $ days) and $\alpha=\pi/3$.
The FIM in the LW approximation can also be written in the compact forms \eqref{gammaijTQ} and \eqref{gammaijTQ2} with
\begin{equation}
\label{lisamdm_ad}
\begin{split}
     M^{dm}&=\left(\frac{2\pi f R}{c} \right)^2\begin{pmatrix}
\textbf{A}_{2\times2}, \textbf{0}_{2\times4}\\
\textbf{0}_{4\times2},\textbf{0}_{4\times4}
\end{pmatrix}, \\
M^{ad}&=\frac{2\pi f R}{c} \begin{pmatrix}
\textbf{B}_{2\times2}, \textbf{C}_{2\times4}\\
\textbf{C}^T_{4\times2},\textbf{0}_{4\times4}
\end{pmatrix},
\end{split}
\end{equation}
where the matrix elements of $M^{am}$, $M^{dm}$ and $M^{ad}$ are presented in Appendix \ref{AppendixLISA}.
The matrix $M^{am}$ is non-singular and independent of the frequency.
In the matrix $M^{dm}$,
only the submatrix $\bf{A}$ is nonzero.

From Eqs. \eqref{gamma} and \eqref{gammaijTQ2}, we see that the FIM is proportional to $\mathcal{A}^2/S_n(f)$,
so $\Delta \Omega\propto S_n(f)/\mathcal{A}^2$ and the parameter estimation errors are $\sigma_{i}\propto \sqrt{S_n(f)}/\mathcal{A}$.
For convenience, we write the parameter estimation errors as
\begin{equation}
\Delta\Omega=\frac{S_n(f)}{\mathcal{A}^2}\Delta\tilde\Omega,\qquad \sigma_i=\frac{\sqrt{S_n(f)}}{\mathcal{A}}\tilde\sigma_i,
\end{equation}
to factor out the effect of the noise curve $S_n(f)$ because it is just a number for a monochromatic source.

\begin{figure}[htp]
\centering
  \includegraphics[width=0.8\columnwidth]{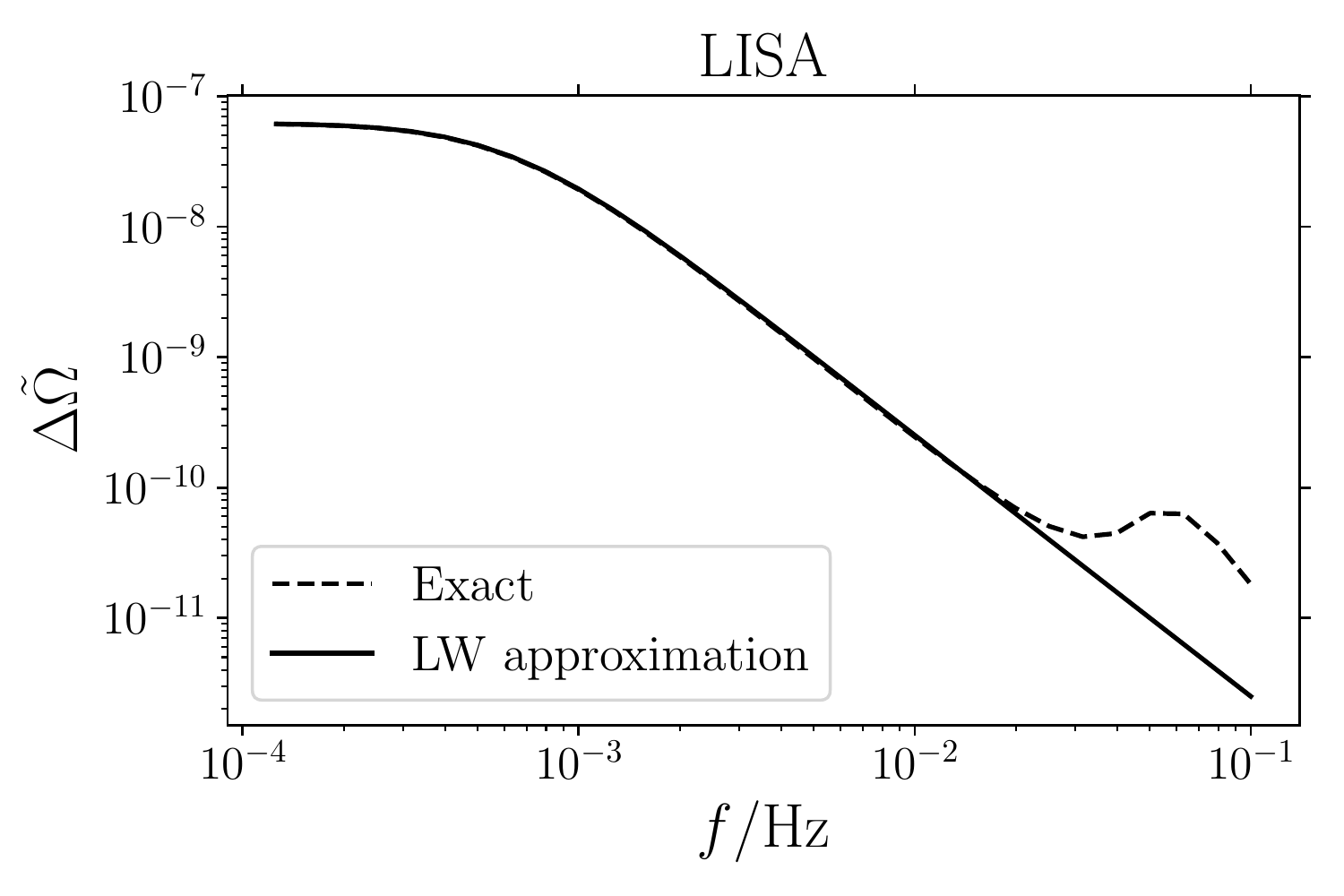}
\\
  \includegraphics[width=0.8\columnwidth]{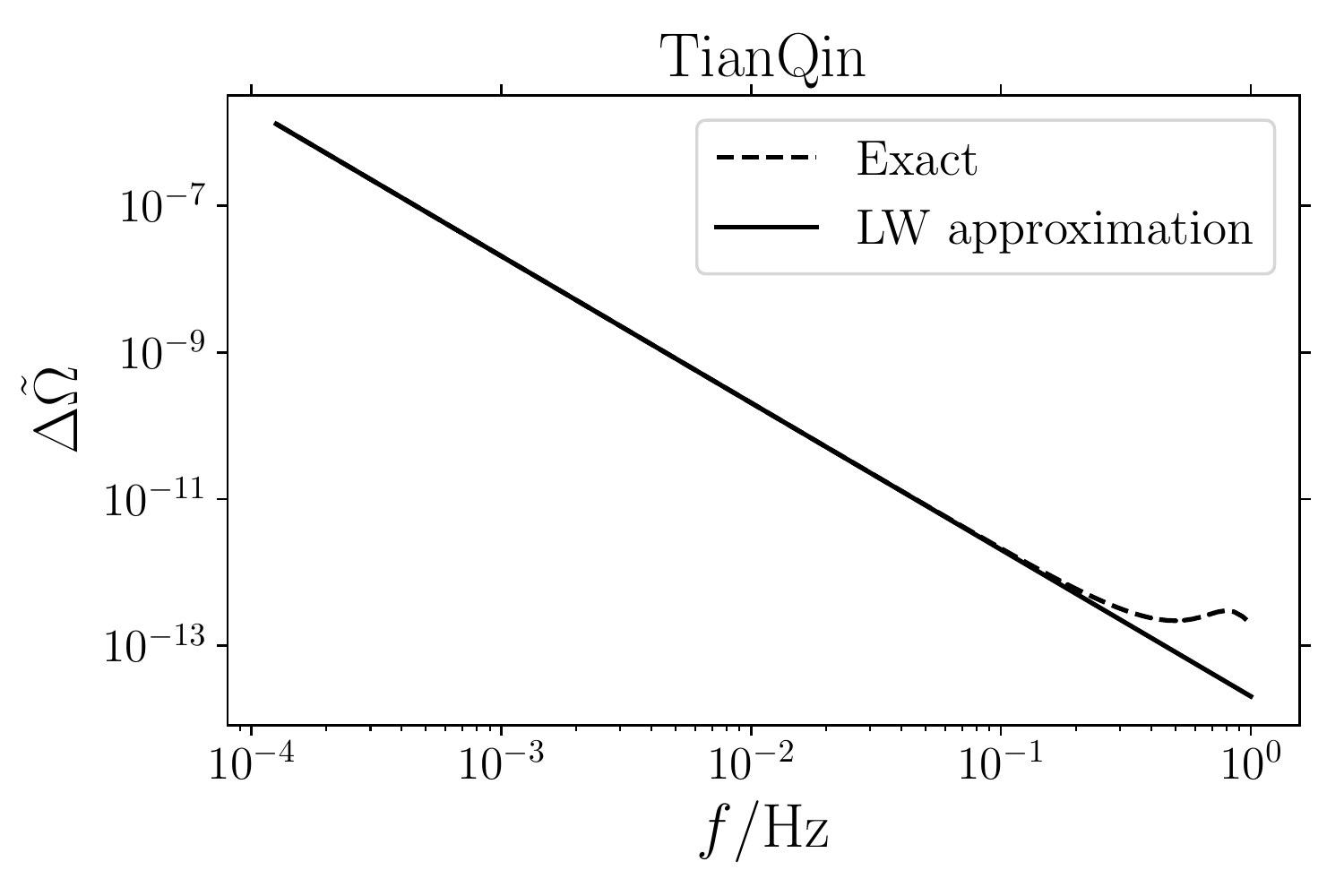}
  \caption{The normalized angular resolutions $\Delta \tilde \Omega$ of LISA and TianQin as functions of frequency for the source with ($\theta=\pi/5, \phi=4\pi/3, \iota=\pi/4, \psi=\pi/4, \phi_0=0 $).
  The dashed curve represents the exact $\Delta \tilde\Omega$ without the LW approximation, the solid curve represents the $\Delta \tilde\Omega$ calculated with the LW approximation.
  }
  \label{tqlisa1}
\end{figure}

Using the FIM derived above, we get the normalized angular resolutions $\Delta \tilde\Omega$ of LISA and TianQin as shown in Fig. \ref{tqlisa1}.
In Fig. \ref{tqlisa1}, we compare the normalized angular resolutions $\Delta \tilde\Omega$ of LISA and TianQin with and without the LW approximation.
From Fig. \ref{tqlisa1}, we see that the approximate result of $\Delta \tilde\Omega$ derived with the LW approximation is almost the same as the  exact result derived without the LW approximation when the GW frequency is less than the transfer frequency,
i.e., $f\leq 0.02$ Hz for LISA and $f\leq 0.28$ Hz for TianQin, respectively.
At higher frequencies, $f>f^*$, the transfer function
$T(f, \hat{u}\cdot\hat{w})$ in Eq. (\ref{transferfunction}) is no longer constant and becomes frequency dependent.
In other words, the arm length of the detector is shorter than
the GW's wavelength and the LW approximation breaks down when the frequency $f>f^*$.
In summary, the LW approximation works quite well when the frequencies are below $10^{-2}$ Hz for LISA and TianQin,
so we can use the LW approximation to estimate the errors of parameters in the low ($\sim10^{-4}$ Hz) and medium ($\sim10^{-2}$ Hz) frequency regimes.

As seen from Fig. \ref{tqlisa1}, for LISA, the normalized angular resolution $\Delta \tilde\Omega$ is almost independent of the frequency below sub-mHz and depends on the frequency in a power-law form till $f\sim f^*_s$.
However, the normalized angular resolution $\Delta \tilde\Omega$ of TianQin depends on the frequency in a power-law form when $f< f^*_t$.
To understand these results, we use the inverse of the FIM, $(\Gamma^{LW})^{-1}$
to estimate the parameter errors.
In general, it is not easy to obtain the analytical expression for $(\Gamma^{LW})^{-1}$.
In Appendix \ref{InverseFIM},
we give the frequency dependent behaviour of $(\Gamma^{LW})^{-1}$ in the low and high frequency limits so that we can understand how the amplitude and Doppler phase modulations contribute to the parameter error estimations.
Note that the frequency limit (especially the high frequency limit) is relative and it still satisfies the LW condition.
The main results obtained in Appendix \ref{InverseFIM} are as follows.
For TianQin in the low frequency limit $f\rightarrow 0$,
the errors $\tilde\sigma_{i}$ fall off as $1/f$ because $M^{am}$ is singular
and the contribution is from the Doppler phase modulation only.
In the  high frequency limit $f\rightarrow\infty$, the errors $\tilde\sigma_{i}$ fall off as $1/f$ for $i=(\theta, \phi)$ and approach a constant for $i=(\ln{\mathcal{A}},\iota,\psi,\phi_0)$.
For LISA in the low frequency limit $f\rightarrow 0$,
the errors $\tilde\sigma_{i}$ approach a constant because $M^{am}$ is non-singular
and the amplitude modulation dominates over the Doppler phase modulation.
In the  high frequency limit $f\rightarrow\infty$, the errors $\tilde\sigma_{i}$ fall off as $1/f$ for $i=(\theta, \phi)$ and approach a constant for $i=(\ln{\mathcal{A}},\iota,\psi,\phi_0)$.
These results are consistent with Fig. \ref{tqlisa1}.

\begin{figure}
  \centering
  \includegraphics[width=0.8\columnwidth]{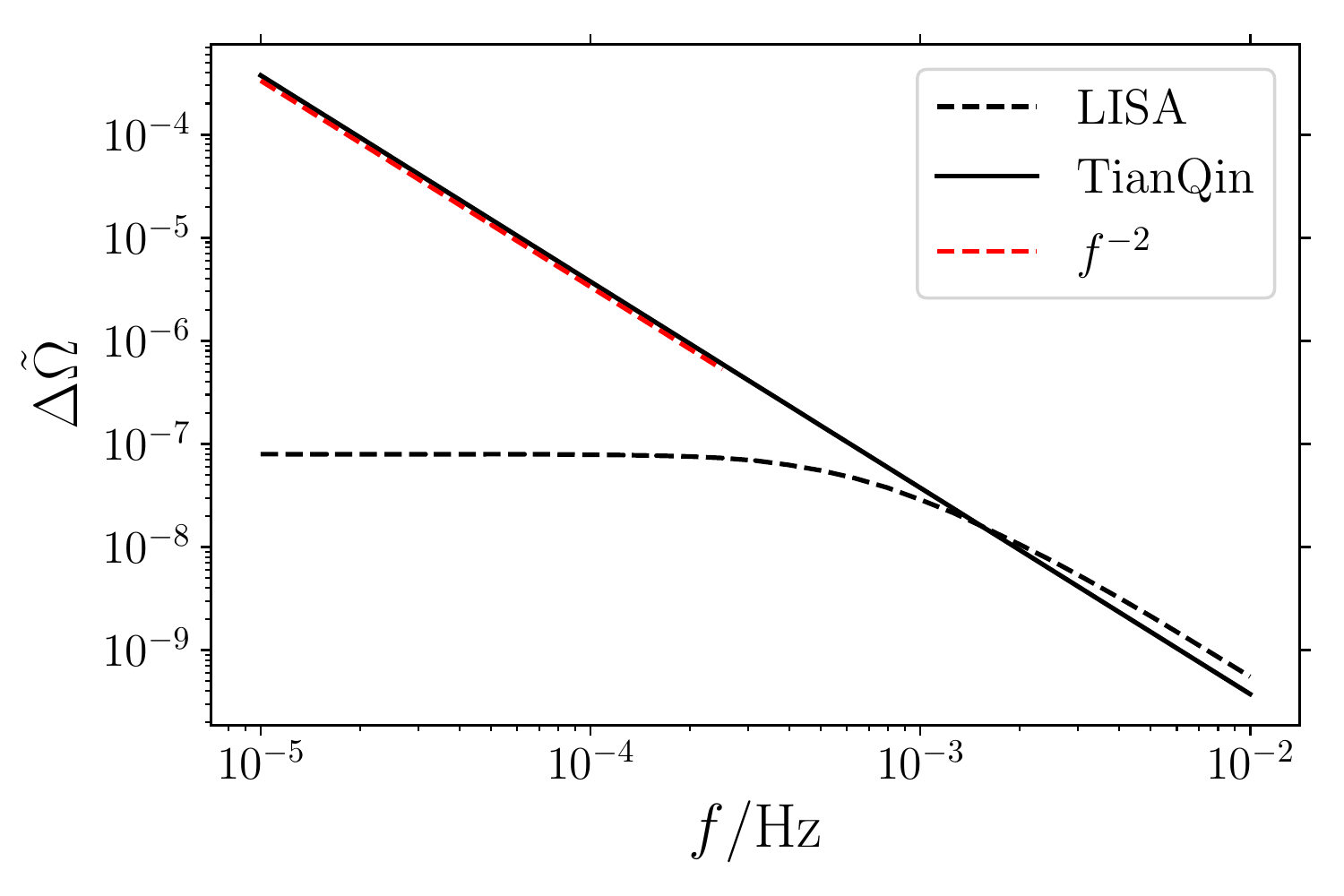}
  \caption{The medians of the normalized angular resolutions $\Delta \tilde\Omega$
  of LISA and TianQin
  as functions of frequency for monochromatic sources.}\label{omegas2}
\end{figure}

\begin{figure*}[htp]
  \centering
  \includegraphics[width=0.9\textwidth]{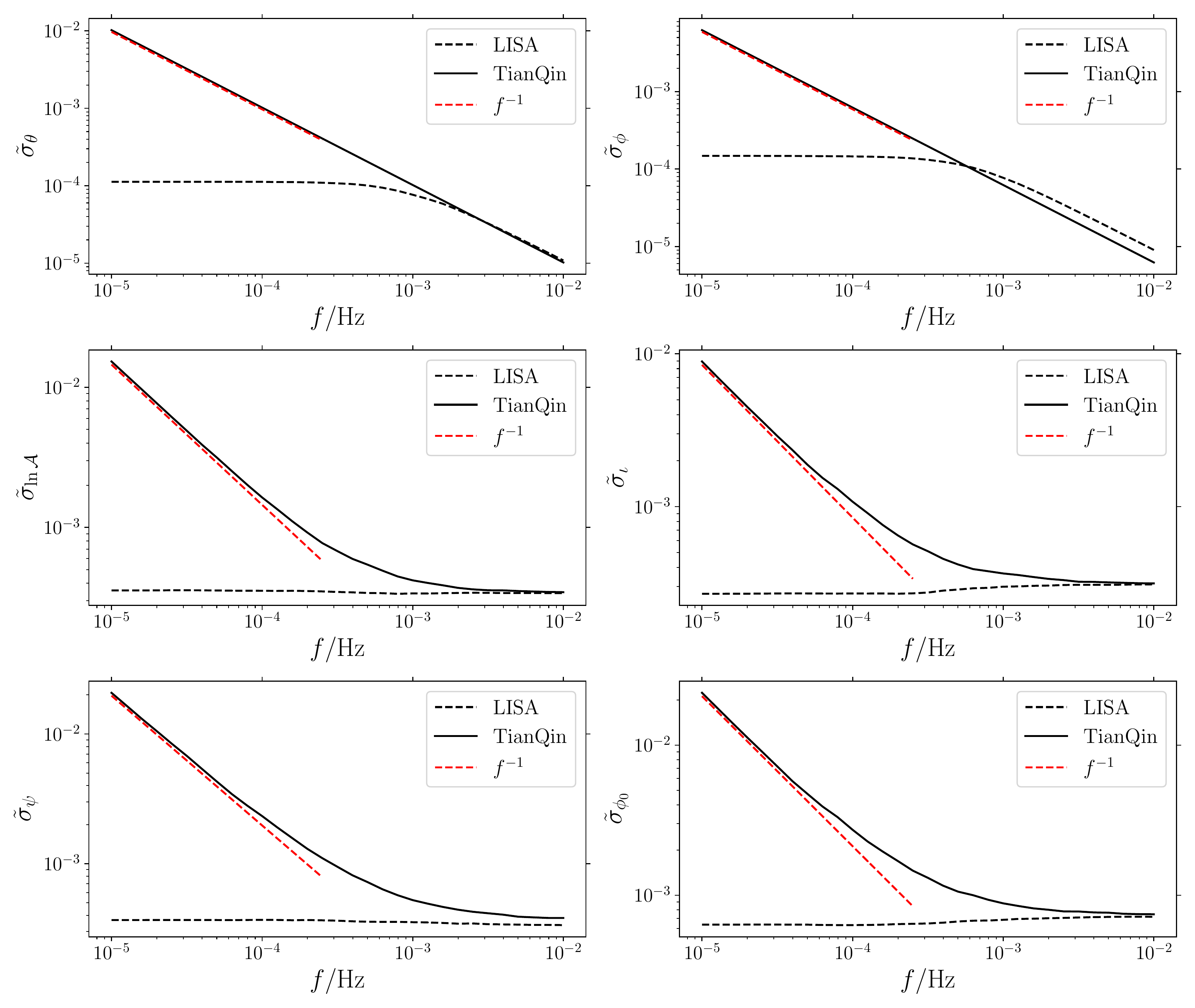}
  \caption{The medians of parameter estimation errors of LISA and TianQin as functions of frequency for monochromatic sources in the low and medium frequency regimes.}
  \label{parametersLWaverage}
\end{figure*}

To show the frequency dependent behavior is general and independent of the particular choice of the source,
we simulate 2000 sources with parameters ($\cos\theta,\phi,\cos\iota,\psi,\phi_0$) uniformly distributed.
The medians of parameter estimation errors are shown in Figs. \ref{omegas2} and \ref{parametersLWaverage}.
Fig. \ref{omegas2} shows the normalized angular resolutions $\Delta \tilde\Omega$ of LISA and TianQin.
As expected, they behave differently in the low frequency regimes.
The angular resolution $\Delta\Omega$ of LISA  depends on the frequency as $S_n(f)$ but the angular resolution $\Delta\Omega$ of TianQin depends on the frequency as $S_n(f)/f^{2}$.
For LISA, the matrix $M^{am}$ is non-singular
and independent of the frequency,
so it contributes to the parameter estimation and it helps the localization of the source due to the changing orientation of the detector plane.
As the frequency decreases,
both the Doppler phase modulation $\Gamma^{dm}$ which depends on $f^2$ and $\Gamma^{ad}$ which depends on $f$ tend to be 0 compared with the total amplitude modulation $\Gamma^{am}$ which is independent of $f$,
so the normalized angular resolution $\Delta\tilde\Omega$ of LISA approaches to a constant.
For TianQin, the total amplitude modulation $\Gamma^{am}$ is singular,
so its ability of localization comes only from the Doppler phase modulation $\Gamma^{dm}$ which depends on the frequency as $f^2$.
This means that even though the total amplitude modulation $\Gamma^{am}$ is singular,
but the full Fisher matrix $\Gamma^{LW}$ including the Doppler phase modulation $\Gamma^{dm}$  is non-singular,
so TianQin uses the Doppler phase modulation only to locate the source.
Therefore, the angular resolution $\Delta\Omega$ of TianQin depends on the frequency as $S_n(f)/f^{2}$.
As the frequency $f$ increases,
the main contribution to sky localizations for LISA and TianQin comes from the Doppler phase modulation effect $\Gamma^{dm}$ and their angular resolutions $\Delta\Omega$  have the same behavior of $S_n(f)/f^{2}$.

Fig. \ref{parametersLWaverage} shows the parameter estimation errors $\tilde\sigma_{\boldsymbol{\xi}}$ for LISA and TianQin.
At low frequency regime (below sub-mHz) $\sigma_{\boldsymbol{\xi}}$ is proportional to  $\sqrt{S_n(f)}/f$ for TianQin but $\sigma_{\boldsymbol{\xi}}$  approaches $\sqrt{S_n(f)}$ for LISA.
For TianQin, the Doppler phase modulation $\Gamma^{dm}$ dominates and it improves both the sky localization and the parameter estimation as the frequency increases.
For LISA, as analyzed above, the amplitude modulation  $\Gamma^{am}$ dominates,
so the parameter estimation errors $\tilde\sigma_{\boldsymbol{\xi}}$ are independent of frequency.
In the median frequency regime, 
the errors of the parameters ($\theta, \phi$) are proportional to  $\sqrt{S_n(f)}/f$
and the errors of the parameters ($\ln \mathcal{A},\iota, \psi, \phi_0$) approach to $\sqrt{S_n(f)}$ for both TianQin and LISA.
The difference between the parameters ($\ln\mathcal{A},\iota, \psi, \phi_0$) and the parameters ($\theta, \phi$) is because the $f^2$ dependent Doppler modulation $\Gamma^{dm}$ contributes only to the ($\theta, \phi$) components in the FIM and it dominates the total matrix $\Gamma^{LW}$ in the medium frequency regime.
For the parameters ($\ln\mathcal{A},\iota, \psi, \phi_0$),
the errors  approach to constants because they are determined by the $4\times 4$ submatrix in the total amplitude modulation $\Gamma^{am}$.

If the parameters $\psi$ and $\phi_0$ are known,
then we are left with four parameters $\bm{\xi}=\{\theta,\phi,\ln\mathcal{A},\iota\}$.
In Appendix \ref{TQ4paras}, we give the estimation errors of the four parameters.
In this case, the matrices $\Gamma^{am}$ for LISA and TianQin are both non-singular
and independent of frequency,
so the behaviours of the parameter errors for LISA and TianQin are similar
except that the parameter errors for LISA are much smaller in the low frequency regime.
We also simulate the signal in LISA and use Bayesian analysis to estimate the parameter errors and the results are presented in Appendix \ref{bayes1}.
The results show that the parameter estimation errors with the Bayesian analysis 
are $3-6$ times larger that those with the FIM.

\subsection{The high frequency regime}

\begin{figure*}[htp]
  \centering
  \includegraphics[width=0.9\textwidth]{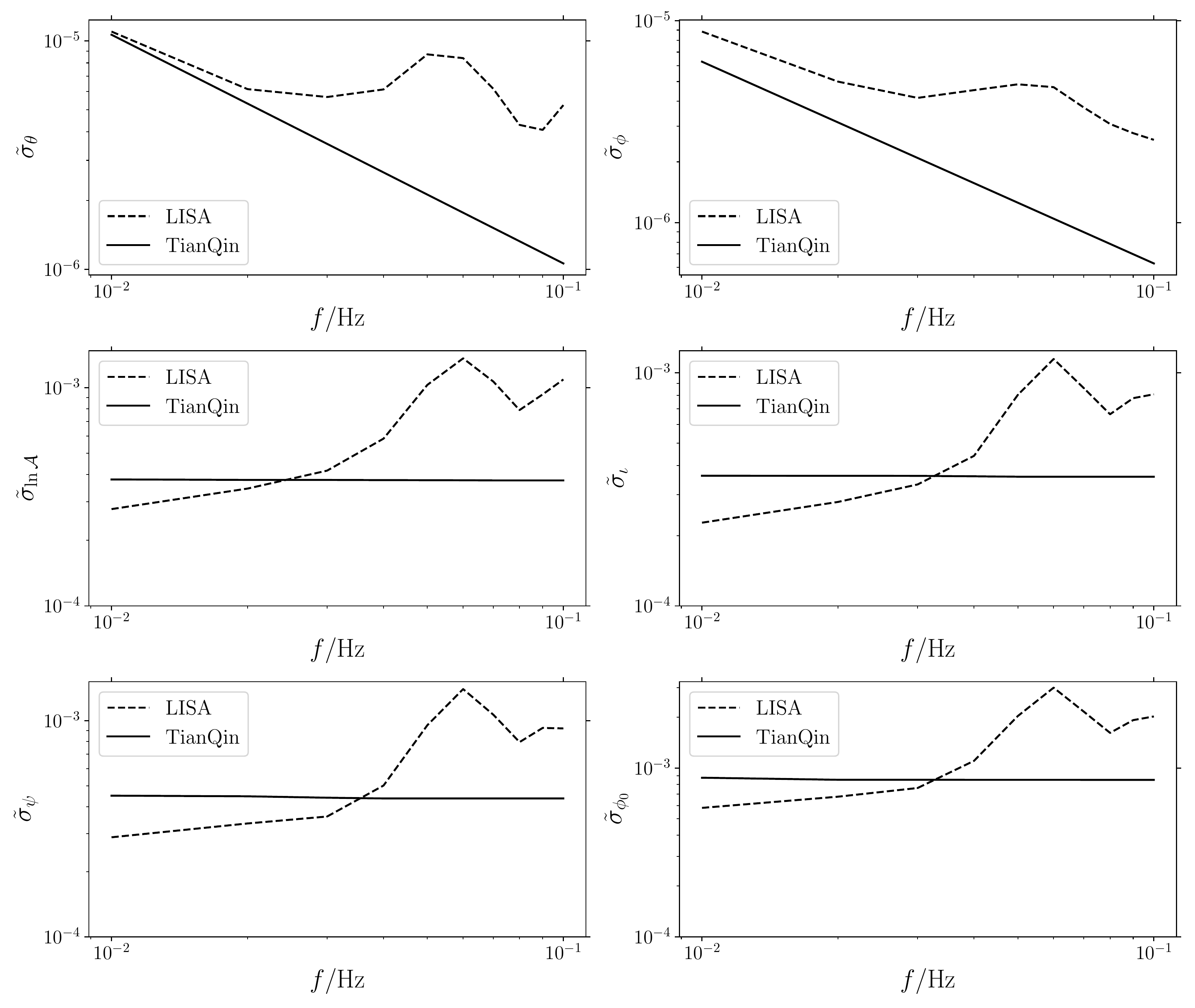}
  \caption{The medians of parameter estimation errors of LISA and TianQin as functions of frequency for monochromatic sources in the high frequency regimes.}
  \label{hparameters}
\end{figure*}

In the high frequency regime from $10^{-2}$ Hz to $10^{-1}$ Hz,
the LW approximation breaks down for LISA.
We numerically calculate the parameter errors for LISA.
Fig. \ref{hparameters} shows the results of $\tilde\sigma_{\boldsymbol{\xi}}$ for LISA and TianQin at high frequencies.
When the frequency $f>0.02$ Hz, we see from Fig. \ref{hparameters} that  the errors of parameters become larger for LISA.
The reason is that the transfer function $T(f, \hat{u}\cdot\hat{w})$ in Eq. (\ref{transferfunction})  becomes less than 1 and decreases as the frequency $f$ increases.
However, at the same frequency the transfer function of TianQin is much bigger than that of LISA,
so the estimation errors of parameters for TianQin are smaller than LISA.

\section{Conclusion}
After splitting the FIM into three parts,
we separately analyze the effects on the parameter estimation errors by the amplitude modulation due to the changing orientation of the detector plane and the Doppler phase modulation due to the translational motion of the center of the detector around the Sun.
In the low and medium frequency regimes,
we take the long-wavelength approximation to give analytical formulas for the parameter estimation errors.
We find that in the low frequency regime,
the angular resolution  falls off as $S_n(f)/f^2$ for TianQin but $S_n(f)$ for LISA, and the estimation errors of the other parameters fall off as $\sqrt{S_n(f)}/f$ for TianQin but $\sqrt{S_n(f)}$ for LISA.
The different pattern between LISA and TianQin is because the total amplitude modulation is missing in TianQin.
For LISA the total amplitude modulation
dominates over the Doppler phase modulation,
so the parameter error $\tilde{\sigma}_i$
is independent of frequency;
For TianQin only the Doppler phase modulation contributes to the parameter estimation, so the parameter error $\tilde{\sigma}_i$ falls off as $1/f$.
In the medium frequency regime,
the Doppler phase modulation dominates and
it affects on the angular resolutions only,
but its effect on the other parameters is negligible.
Because of this,
for both TianQin and LISA,
the angular resolutions fall off as $S_n(f)/f^2$,
and the estimation errors of the parameters ($\ln\mathcal{A},\iota, \psi, \phi_0 $) fall off as $\sqrt{S_n(f)}$
because they are determined by the frequency independent $4\times 4$ submatrix in the total amplitude modulation $\Gamma^{am}$.
These results are also confirmed by numerical method.
In the high frequency regime where the long-wavelength approximation fails,
we numerically calculate the parameter  estimation errors for LISA and TianQin and find that because of different transfer frequency due to the difference in arm lengths,
the parameter estimation errors measured by TianQin are smaller than LISA.
Although the parameter estimations with the Bayesian analysis are more reliable and
the estimation errors are $3-6$ times larger that those with the FIM,
the FIM helps us understand the effects on the parameter estimations by the total amplitude modulation and the Doppler phase modulation.
The results are useful for understanding the parameter estimation errors measured by LISA and TianQin due to the difference in the constellation.

\begin{acknowledgments}
This research is supported in part by the National Key Research and Development Program of China under Grant No. 2020YFC2201504;
the National Natural Science
Foundation of China under Grant Nos. 11875136 and 12075202;
and the Major Program of the National Natural Science Foundation of China under Grant No. 11690021.
\end{acknowledgments}

\appendix
\section{The FIM for TianQin}\label{AppendixTianQin}

This appendix provides the nonzero matrix elements of FIM for TianQin.
The nonzero elements of the matrix $M^{am}$ for TianQin in the detector coordinate system are
\begin{equation}
\label{TianQinamplitude1}
\nonumber
\begin{split}
  M^{am}_{11}    &  =  \frac{3}{32}   \sin ^2(\theta_d) \left(-16 \sin ^2(\theta_d) \sin ^4(\iota ) \cos (4 \psi_d )\right.\\
 &\left.+(\cos (2 \theta_d)+3) (28 \cos (2 \iota )+\cos (4 \iota )+35)\right),    \\
 M^{am}_{12}    &  = -\frac{3}{2}  \sin ^3(\theta_d) \sin ^4(\iota ) \sin (4 \psi ),   \\
 M^{am}_{13}    &  =   \frac{3}{128}   \sin (2 \theta_d) \left(16 \sin ^2(\theta_d) \sin ^4(\iota ) \cos (4 \psi_d )\right.\\
&\left.-(\cos (2 \theta_d)+7) (28 \cos (2 \iota )+\cos (4 \iota )+35)\right),       \\
 M^{am}_{14}    &  =   \frac{3}{32}   \sin (2 \theta_d) \sin (2 \iota ) \left(4 \sin ^2(\theta_d) \sin ^2(\iota ) \cos (4 \psi_d )\right.\\
 &\left.+(\cos (2 \theta_d)+7) (\cos (2 \iota )+7)\right),       \\
 M^{am}_{15}    &  =    - \frac{3}{2}    \sin ^3(\theta_d) \cos (\theta_d) \sin ^4(\iota ) \sin (4 \psi_d ),     \\
 M^{am}_{22}    &  =  \frac{3}{128}   \left(64 \sin ^4(\theta_d) \sin ^4(\iota ) \cos (4 \psi_d )\right.\\
&\left.+(28 \cos (2 \theta_d)+\cos (4 \theta_d)+35)\right.\\
&\left. \times(28 \cos (2 \iota )+\cos (4 \iota )+35)\right),        \\
 M^{am}_{25}    &  =  \frac{3}{16}   (7 \cos (\theta_d)+\cos (3 \theta_d))\\
 &\times(28 \cos (2 \iota )+\cos (4 \iota )+35),        \\
 \end{split}
\end{equation}

\begin{equation}
\nonumber
\begin{split}
 M^{am}_{26}    &  =  -  \frac{3 }{4}   (7 \cos (\theta_d)+\cos (3 \theta_d)) (7 \cos (\iota )+\cos (3 \iota )),      \\
 M^{am}_{33}    &  = \frac{3}{512} \left(64 \sin ^4(\theta_d) \sin ^4(\iota ) \cos (4 \psi_d )\right.\\
&\left.+(28 \cos (2 \theta_d)+\cos (4 \theta_d)+35)\right.\\
&\left.\times(28 \cos (2 \iota )+\cos (4 \iota )+35)\right),         \\
 M^{am}_{34}    &  =   - \frac{3}{512}  \sin (2 \iota ) \left(-64 \sin ^4(\theta_d) \sin ^2(\iota ) \cos (4 \psi_d ) \right.\\
&\left.+4 (28 \cos (2 \theta_d)+\cos (4 \theta_d)+35) (\cos (2 \iota )+7)\right),      \\
 M^{am}_{35}    & =-\frac{3}{4}   \sin ^4(\theta_d) \sin ^4(\iota ) \sin (4 \psi_d ), \\
\end{split}
\end{equation}

\begin{equation}\label{TianQinamplitude2}
\nonumber
\begin{split}
 M^{am}_{44}    &  =   \frac{3}{32}   \sin ^2(\iota ) \left(-16 \sin ^4(\theta_d) \sin ^2(\iota ) \cos (4 \psi_d )\right.\\
&\left.+(28 \cos (2 \theta_d)+\cos (4 \theta_d)+35) (\cos (2 \iota )+3)\right),     \\
 M^{am}_{45}    & =  - \frac{3}{2}    \sin ^4(\theta_d) \sin ^3(\iota ) \cos (\iota ) \sin (4 \psi_d )     ,  \\
 M^{am}_{46}    & =\frac{3}{4}   \sin ^4(\theta_d) \sin ^3(\iota ) \sin (4 \psi_d ),          \\
 M^{am}_{55}    & =  \frac{3}{128}   \left( -64 \sin ^4(\theta_d) \sin ^4(\iota ) \cos (4 \psi_d )  \right.\\
&\left. (28 \cos (2 \theta_d)+\cos (4 \theta_d)+35)\right.\\
&\left.\times(28 \cos (2 \iota )+\cos (4 \iota )+35)\right),        \\
 M^{am}_{56}    & = -\frac{3}{32}   (28 \cos (2 \theta_d)+\cos (4 \theta_d)+35)\\
 &\times(7 \cos (\iota )+\cos (3 \iota )),         \\
 M^{am}_{66}    & = \frac{3}{512}   \left(64 \sin ^4(\theta_d) \sin ^4(\iota ) \cos (4 \psi_d )\right.\\
&\left.+(28 \cos (2 \theta_d)+\cos (4 \theta_d)+35)\right.\\
&\left.\times(28 \cos (2 \iota )+\cos (4 \iota )+35)\right).         \\
\end{split}
\end{equation}

It is straightforward to check that $\mathrm{det}~ M^{am}=0$. 
For the same source in the sky, 
the parameters $(\theta_d,\phi_d,\psi_d)$ in the detector coordinate system  are related with the parameters $(\theta,\phi,\psi)$ in the heliocentric coordinate system as
\begin{equation*}
    \begin{split}
\theta_d &=\arccos (\sin (\alpha ) \sin (\beta ) \sin (\phi) \sin (\theta)\\
&+\cos (\alpha ) \sin (\beta ) \cos (\phi) \sin (\theta)+\cos (\beta ) \cos (\theta)),\\
\phi_d&=\arctan \left(    \right.\\
&\left. (\cos (\alpha ) \sin (\phi) \sin (\theta)-\sin (\alpha ) \cos (\phi) \sin (\theta))  \right.\\
&\left. /(\cos (\alpha ) \cos (\beta ) \cos (\phi) \sin (\theta)  \right.\\
&\left. +\sin (\alpha ) \cos (\beta ) \sin (\phi) \sin (\theta)-\sin (\beta ) \cos (\theta))  \right),\\
\psi_d&=\arctan \left( \left( \cos (\beta ) \sin (\theta) \sin (\psi ) +\sin (\beta )\cos (\alpha )\times \right.\right. \\
&\left(   (\sin (\phi) (-\cos (\psi ))-\cos (\phi) \cos (\theta) \sin (\psi ))\right.\\
&\left.\left.+\tan (\alpha )(\cos (\phi) \cos (\psi )-\sin (\phi) \cos (\theta) \sin (\psi ))              \right)\right)  \\
&/\left( \sin (\alpha ) \sin (\beta ) (\sin (\phi) \cos (\theta) \cos (\psi )+\cos (\phi) \sin (\psi )) \right.\\
&\left. +\cos (\alpha ) \sin (\beta ) (\cos (\phi) \cos (\theta) \cos (\psi )-\sin (\phi) \sin (\psi ))\right.\\
&\left.\left.-\cos (\beta ) \sin (\theta) \cos (\psi )    \right)  \right).\\
\end{split}
\end{equation*}
The matrix $ M^{am}$ for TianQin in the heliocentric coordinate system becomes
\begin{equation*}
 M^{am}_{ij} \rightarrow F^{li} M^{am}_{lk} F^{kj},
\end{equation*}
where $F^{ij}=\partial\boldsymbol{\hat{\xi}}_i/\partial\boldsymbol{\xi}_j$ and $\boldsymbol{\hat{\xi}}=(\theta_d, \phi_d, \ln\mathcal{A},\iota, \psi_d, \phi_0$).

The nonzero elements of the matrix $ M^{dm}$ for TianQin in the heliocentric coordinate system are
\begin{equation}
\label{TianQindoppler}
\begin{split}
 M^{dm}_{11}    & = \frac{2 \pi ^2\rho^2 f^2 R^2}{c^2}     \cos ^2(\theta), \\
 M^{dm}_{22}    & = \frac{2 \pi ^2\rho^2 f^2 R^2}{c^2}   \sin ^2(\theta),
 \end{split}
\end{equation}
where

\begin{equation}
\nonumber
\begin{split}
\rho^2&=\frac{3}{512}   \left( 64 \sin ^4(\theta_d) \cos (4 \psi_d) \sin ^4(\iota )   \right.\\
&\left.+  (28 \cos (2 \theta_d)+\cos (4 \theta_d)+35)\right.\\
&\left.\times(28 \cos (2 \iota )+\cos (4 \iota )+35) \right).
\end{split}
\end{equation}

\section{The FIM for LISA}
\label{AppendixLISA}

This appendix provides the nonzero matrix elements of FIM for LISA in the heliocentric coordinate system.
\begin{widetext}
\begin{equation}
\label{LISAamplitude1}
\nonumber
\begin{split}
 M^{am}_{11} &=\frac{  \sin ^2(\alpha )}{16384}
    \left(   -20736 \sin ^2(\theta ) \cos (\theta ) \sin ^4(\iota ) \sin (4 \psi ) \sin (2 \alpha -4 \phi )          \right.\\
    &+1296 (28 \cos (2 \iota )+\cos (4 \iota )+35)  \sin ^4(\theta ) \cos (2 \alpha -4 \phi )                                   \\
    &+4 (28 \cos (2 \iota )+\cos (4 \iota )+35)(76 \cos (2 \theta )+37 \cos (4 \theta )+847)                                \\
    &\left.        -64 \sin ^2(\theta ) \sin ^4(\iota ) \cos (4 \psi ) (81 (\cos (2 \theta )+3) \cos (2 \alpha -4 \phi )+74 \cos (2 \theta )-2)           \right),  \\
 M^{am}_{12}&=\frac{  \sin ^2(\alpha ) }{1024}
\left(\sin ^4(\iota ) \left(81 (14 \sin (2 \theta )+\sin (4 \theta )) \cos (4 \psi ) \sin (2 \alpha -4 \phi )\right.\right.\\
&               -81 \sin ^3(\theta ) \cos (\theta ) (28 \cos (2 \iota )+\cos (4 \iota )+35) \sin (2 \alpha -4 \phi )\\
&                           -324 \sin ^4(\iota )\sin (\theta ) \sin (4 \psi ) \cos (2 \alpha ) (3 \cos (2 \theta )+5) \cos (4 \phi )                                       \\
&   \left.                        -324 \sin ^4(\iota )\sin (\theta ) \sin (4 \psi )\sin (2 \alpha ) (3 \cos (2 \theta )+5) \sin (4 \phi )-76 \sin ^2(\theta )     \right),      \\
 M^{am}_{13}&=\frac{  \sin ^2(\alpha )}{16384} \left(1296 (7 \sin (\theta )+3 \sin (3 \theta )) \sin ^4(\iota ) \sin (4 \psi ) \sin (2 \alpha -4 \phi )\right.\\
&+1296\sin (2 \theta )  \sin ^4(\iota ) \cos (4 \psi )  \cos (2 \alpha ) (\cos (2 \theta )+7) \cos (4 \phi )    \\
&+648 \sin (2 \theta )       2 \sin ^4(\iota ) \cos (4 \psi ) \sin (2 \alpha ) (\cos (2 \theta )+7) \sin (4 \phi )-296 \sin ^2(\theta )                                                                        \\
&\left.\left.+(28 \cos (2 \iota )+\cos (4 \iota )+35) \left(-81 \sin ^2(\theta ) \cos (2 \alpha -4 \phi )+37 \cos (2 \theta )+115\right)\right)\right),\\
 M^{am}_{14}&=\frac{  \sin ^2(\alpha ) }{8192}\left(1296 (7 \sin (\theta )+3 \sin (3 \theta )) \sin ^3(\iota ) \cos (\iota ) \sin (4 \psi ) \sin (2 \alpha -4 \phi )\right.\\
&+1296\sin (2 \theta ) \sin ^3(\iota ) \cos (\iota ) \cos (4 \psi )  \cos (2 \alpha ) (\cos (2 \theta )+7) \cos (4 \phi )\\
&       -4 \sin (2 \theta ) (14 \sin (2 \iota )+\sin (4 \iota )) \left(-81 \sin ^2(\theta ) \cos (2 \alpha -4 \phi )+37 \cos (2 \theta )+115\right)      \\
&\left.       +8 \sin (2 \theta ) \sin ^3(\iota ) \cos (\iota ) \cos (4 \psi )\left(162 \sin (2 \alpha ) (\cos (2 \theta )+7) \sin (4 \phi )-296 \sin ^2(\theta )\right)                  \right),\\
\end{split}
\end{equation}

\begin{equation}
\nonumber
\begin{split}
 M^{am}_{15}&=\frac{  \sin ^2(\alpha ) }{1024}\left(162 (7 \sin (\theta )+3 \sin (3 \theta )) \sin ^4(\iota ) \cos (4 \psi ) \sin (2 \alpha -4 \phi )\right.\\
& -81 \sin ^3(\theta ) (28 \cos (2 \iota )+\cos (4 \iota )+35) \sin (2 \alpha -4 \phi )\\
&-162 \cos (2 \alpha ) (\cos (2 \theta )+7) \cos (4 \phi )\sin (2 \theta ) \sin ^4(\iota ) \sin (4 \psi ) \\
& \left.   -\sin (2 \theta ) \sin ^4(\iota ) \sin (4 \psi )\left(162 \sin (2 \alpha ) (\cos (2 \theta )+7) \sin (4 \phi )-296 \sin ^2(\theta )\right)    \right),\\
 M^{am}_{16}    & =  \frac{81}{256}   \sin ^2(\alpha ) \sin ^3(\theta ) (7 \cos (\iota )+\cos (3 \iota )) \sin (2 \alpha -4 \phi ),       
\end{split}
\end{equation}

\begin{equation}
\label{LISAamplitude2}
\begin{split}
 M^{am}_{22}&=\frac{  \sin ^2(\alpha ) }{16384}\left(8 \sin ^4(\iota ) \cos (4 \psi ) \left(162 \cos (2 \alpha ) (28 \cos (2 \theta )+\cos (4 \theta )+35) \cos (4 \phi )\right.\right)\\
&+10368 (7 \cos (\theta )+\cos (3 \theta )) \sin ^4(\iota ) \sin (4 \psi ) \sin (2 \alpha -4 \phi )\\
&+4 (28 \cos (2 \iota )+\cos (4 \iota )+35)(1735-19 \cos (4 \theta )) \\
&+4 (28 \cos (2 \iota )+\cos (4 \iota )+35) \left(324 \sin ^4(\theta ) \cos (2 \alpha -4 \phi )+908 \cos (2 \theta )\right)\\
&\left.   +8 \sin ^4(\iota ) \cos (4 \psi )\left(162 \sin (2 \alpha ) (28 \cos (2 \theta )+\cos (4 \theta )+35) \sin (4 \phi )-608 \sin ^4(\theta )\right)     \right),\\
 M^{am}_{23}&=\frac{81   \sin ^2(\alpha ) }{2048}\left(8\sin ^4(\iota )  (7 \cos (\theta )+\cos (3 \theta )) \sin (4 \psi ) \cos (2 \alpha -4 \phi )\right.\\
&-\sin ^4(\theta ) (28 \cos (2 \iota )+\cos (4 \iota )+35) \sin (2 \alpha -4 \phi )\\
&\left.-\sin ^4(\iota )(28 \cos (2 \theta )+\cos (4 \theta )+35) \cos (4 \psi ) \sin (2 \alpha -4 \phi )\right),\\
 M^{am}_{24}&=\frac{81   \sin ^2(\alpha ) \sin (\iota ) \cos (\iota ) }{1024}\left(8 (7 \cos (\theta )+\cos (3 \theta )) \sin ^2(\iota ) \sin (4 \psi ) \cos (2 \alpha -4 \phi )\right.\\
&+4\sin (2 \alpha -4 \phi )  \sin ^4(\theta ) (\cos (2 \iota )+7)   \\
&\left.-\sin (2 \alpha -4 \phi )\sin ^2(\iota ) \cos (4 \psi ) (28 \cos (2 \theta )+\cos (4 \theta )+35) \right),
\end{split}
\end{equation}

\begin{equation}
\begin{split}
\nonumber
 M^{am}_{25}&=\frac{  \sin ^2(\alpha ) }{1024}\left(648 (7 \cos (\theta )+\cos (3 \theta )) \sin ^4(\iota ) \cos (4 \psi ) \cos (2 \alpha -4 \phi )\right.\\
&\left.+2 (347 \cos (\theta )-19 \cos (3 \theta )) (28 \cos (2 \iota )+\cos (4 \iota )+35)\right.\\
&\left.+81 (28 \cos (2 \theta )+\cos (4 \theta )+35) \sin ^4(\iota ) \sin (4 \psi ) \sin (2 \alpha -4 \phi )\right),\\
 M^{am}_{26}    & =       \frac{1}{128}   \sin ^2(\alpha ) (19 \cos (3 \theta )-347 \cos (\theta )) (7 \cos (\iota )+\cos (3 \iota )),  
 \end{split}
\end{equation}

\begin{equation}
\nonumber
\begin{split}
 M^{am}_{33}&=\frac{\sin ^2(\alpha ) }{65536}
\left(-16 \sin ^4(\iota ) \cos (4 \psi ) \left(81 \cos (2 \alpha ) (28 \cos (2 \theta )+\cos (4 \theta )+35) \cos (4 \phi )\right)\right.\\
&-4 (28 \cos (2 \iota )+\cos (4 \iota )+35)(37 \cos (4 \theta )-3121)\\
&-10368 (7 \cos (\theta )+\cos (3 \theta )) \sin ^4(\iota ) \sin (4 \psi ) \sin (2 \alpha -4 \phi )      \\
&-4 (28 \cos (2 \iota )+\cos (4 \iota )+35) \left(324 \sin ^4(\theta ) \cos (2 \alpha -4 \phi )+460 \cos (2 \theta )\right)\\
&\left. -16 \sin ^4(\iota ) \cos (4 \psi )\left(81 \sin (2 \alpha ) (28 \cos (2 \theta )+\cos (4 \theta )+35) \sin (4 \phi )+592 \sin ^4(\theta )\right)   \right),\\
 M^{am}_{34}&=\frac{  \sin ^2(\alpha ) }{32768}
\left(      8 \sin ^3(\iota ) \cos (\iota )         \left(-1296 (7 \cos (\theta )+\cos (3 \theta )) \sin (4 \psi ) \sin (2 \alpha -4 \phi )\right)  \right.\\
&+4 (14 \sin (2 \iota )+\sin (4 \iota )) \left(324 \sin ^4(\theta ) \cos (2 \alpha -4 \phi )+460 \cos (2 \theta )\right)\\
&-16 \sin ^3(\iota ) \cos (\iota )  \cos (4 \psi ) \left(81 (28 \cos (2 \theta )+\cos (4 \theta )+35) \cos (2 \alpha -4 \phi )\right)\\
&\left.-9472 \sin ^3(\iota ) \cos (\iota )  \cos (4 \psi ) \sin ^4(\theta )+4 (14 \sin (2 \iota )+\sin (4 \iota ))(37 \cos (4 \theta )-3121)\right),\\
 M^{am}_{35}&=\frac{  \sin ^2(\alpha ) \sin ^4(\iota )}{2048} \left(\sin (4 \psi ) \left(81 (28 \cos (2 \theta )+\cos (4 \theta )+35) \cos (2 \alpha -4 \phi )\right)\right.\\
&\left.-648 (7 \cos (\theta )+\cos (3 \theta )) \cos (4 \psi ) \sin (2 \alpha -4 \phi ) +592\sin (4 \psi ) \sin ^4(\theta )\right),\\
\end{split}
\end{equation}

\begin{equation}
\begin{split}
 M^{am}_{44}&=\frac{  \sin ^2(\alpha ) \sin ^2(\iota ) }{4096}
\left(2592 (7 \cos (\theta )+\cos (3 \theta )) \sin ^2(\iota ) \sin (4 \psi ) \sin (2 \alpha -4 \phi )\right.\\
&\left.+4 \sin ^2(\iota ) \cos (4 \psi ) \left(81 \cos (2 \alpha ) (28 \cos (2 \theta )+\cos (4 \theta )+35) \cos (4 \phi )\right.\right.\\
&\left.\left.+81 \sin (2 \alpha ) (28 \cos (2 \theta )+\cos (4 \theta )+35) \sin (4 \phi )+592 \sin ^4(\theta )\right)\right.\\
&\left.-4 (\cos (2 \iota )+3) \left(324 \sin ^4(\theta ) \cos (2 \alpha -4 \phi )+460 \cos (2 \theta )+37 \cos (4 \theta )-3121\right)\right),\\
 M^{am}_{45}&=\frac{  \sin ^2(\alpha ) \sin ^3(\iota ) \cos (\iota ) }{1024}\left(     -648 (7 \cos (\theta )+\cos (3 \theta )) \cos (4 \psi ) \sin (2 \alpha -4 \phi )           \right.\\
&\left.        +\sin (4 \psi ) \left(81 (28 \cos (2 \theta )+\cos (4 \theta )+35) \cos (2 \alpha -4 \phi )+592 \sin ^4(\theta )\right)       \right),\\
 M^{am}_{46}    & =   \frac{  \sin ^2(\alpha ) \sin ^3(\iota ) }{2048}   \left(648 (7 \cos (\theta )+\cos (3 \theta )) \cos (4 \psi ) \sin (2 \alpha -4 \phi )\right.\\
&\left.-\sin (4 \psi ) \left(81 (28 \cos (2 \theta )+\cos (4 \theta )+35) \cos (2 \alpha -4 \phi )+592 \sin ^4(\theta )\right)\right),
\end{split}
\end{equation}

\begin{equation}
\nonumber
\begin{split}
 M^{am}_{55}&=\frac{  \sin ^2(\alpha ) }{16384}
\left(16 \sin ^4(\iota ) \cos (4 \psi ) \left(81 \cos (2 \alpha ) (28 \cos (2 \theta )+\cos (4 \theta )+35) \cos (4 \phi )\right.\right.\\
&-4 (28 \cos (2 \iota )+\cos (4 \iota )+35)(37 \cos (4 \theta )-3121)\\
&\left.+10368 (7 \cos (\theta )+\cos (3 \theta )) \sin ^4(\iota ) \sin (4 \psi ) \sin (2 \alpha -4 \phi )\right.\\
&\left.\left.+81 \sin (2 \alpha ) (28 \cos (2 \theta )+\cos (4 \theta )+35) \sin (4 \phi )+592 \sin ^4(\theta )\right)\right.\\
&\left.-4 (28 \cos (2 \iota )+\cos (4 \iota )+35) \left(324 \sin ^4(\theta ) \cos (2 \alpha -4 \phi )+460 \cos (2 \theta )\right)\right),
\end{split}
\end{equation}

\begin{equation}
\nonumber
\begin{split}
 M^{am}_{56}    & =   \frac{  \sin ^2(\alpha )  }{1024}  (7 \cos (\iota )+\cos (3 \iota )) \left(324 \sin ^4(\theta ) \cos (2 \alpha -4 \phi )\right.\\
&\left.+460 \cos (2 \theta )+37 \cos (4 \theta )-3121\right) , 
\end{split}
\end{equation}

\begin{equation}
\nonumber
\begin{split}
 M^{am}_{66}    & =    \frac{  \sin ^2(\alpha )}{65536}
 \left(-16 \sin ^4(\iota ) \cos (4 \psi ) \left(81 \cos (2 \alpha ) (28 \cos (2 \theta )+\cos (4 \theta )+35) \cos (4 \phi )\right.\right.\\
 &-4 (28 \cos (2 \iota )+\cos (4 \iota )+35)(37 \cos (4 \theta )-3121)\\
 &\left.-10368 (7 \cos (\theta )+\cos (3 \theta )) \sin ^4(\iota ) \sin (4 \psi ) \sin (2 \alpha -4 \phi )\right.\\
  &\left.+81 \sin (2 \alpha ) (28 \cos (2 \theta )+\cos (4 \theta )+35) \sin (4 \phi )+592 \sin ^4(\theta )\right)\\
 &\left.-4 (28 \cos (2 \iota )+\cos (4 \iota )+35) \left(324 \sin ^4(\theta ) \cos (2 \alpha -4 \phi )+460 \cos (2 \theta )\right)\right). 
\end{split}
\end{equation}

\begin{equation}
\label{LISAdoppler}
\begin{split}
   M^{dm}_{11}&=
  \frac{\pi ^2   f^2 R^2 \sin ^2(\alpha ) \cos ^2(\theta )}{32768c^2}
   \left(     8 \sin ^4(\iota ) \cos (4 \psi ) \left(16 \sin ^2(\theta ) (55 \cos (2 \theta )+17)\right)      \right.\\
   &        +  27648 \cos (\theta ) (\cos (2 \theta )-4) \sin ^4(\iota ) \sin (4 \psi ) \sin (2 \alpha -4 \phi )                                                   \\
      &+8 \sin ^4(\iota ) \cos (4 \psi ) \left(54 \cos (2 \alpha ) (-28 \cos (2 \theta )+11 \cos (4 \theta )-175) \cos (4 \phi )\right)\\
   &+432 \sin ^4(\iota ) \cos (4 \psi )\sin (2 \alpha ) (-28 \cos (2 \theta )+11 \cos (4 \theta )-175) \sin (4 \phi )  \\
   &+4 (28 \cos (2 \iota )+\cos (4 \iota )+35) (-1108 \cos (2 \theta )-55 \cos (4 \theta )+3787)\\
   &\left.-216 (28 \cos (2 \iota )+\cos (4 \iota )+35) \sin ^2(\theta ) (11 \cos (2 \theta )+13) \cos (2 \alpha -4 \phi )\right),\\
    M^{dm}_{12}&=\frac{9 \pi ^2   f^2 R^2 \sin ^2(\alpha ) \sin ^3(\theta ) }{4096c^2}
  \left(64 \cos ^2(\theta ) \sin ^4(\iota ) \sin (4 \psi ) (21 \cos (2 \alpha -4 \phi )+2)\right.\\
  &\left.-168 (7 \cos (\theta )+\cos (3 \theta )) \sin ^4(\iota ) \cos (4 \psi ) \sin (2 \alpha -4 \phi )\right.\\
  &\left.-3 (17 \cos (\theta )+7 \cos (3 \theta )) (28 \cos (2 \iota )+\cos (4 \iota )+35) \sin (2 \alpha -4 \phi )\right),
  \end{split}
\end{equation}

\begin{equation}
\nonumber
\begin{split}
   M^{dm}_{22}&=\frac{\pi ^2   f^2 R^2 \sin ^2(\alpha ) \sin ^2(\theta ) }{32768c^2}
  \left(      -8 \sin ^4(\iota ) \cos (4 \psi ) \left(54 \sin (2 \alpha ) (140 \cos (2 \theta )     \right)    \right.\\
  & -6912 (7 \cos (\theta )+5 \cos (3 \theta )) \sin ^4(\iota ) \sin (4 \psi ) \sin (2 \alpha -4 \phi )\\
    &-8 \sin ^4(\iota ) \cos (4 \psi )\left(54 \cos (2 \alpha ) (140 \cos (2 \theta )+17 \cos (4 \theta )+35) \cos (4 \phi )\right)\\
  &-8 \sin ^4(\iota ) \cos (4 \psi ) \left(           +17 \cos (4 \theta )+35) \sin (4 \phi )+16 \sin ^2(\theta ) (91-19 \cos (2 \theta ))                      \right)\\
  &+216 (28 \cos (2 \iota )+\cos (4 \iota )+35) \sin ^2(\theta ) (17 \cos (2 \theta )+7) \cos (2 \alpha -4 \phi )\\
  &\left. +4 (28 \cos (2 \iota )+\cos (4 \iota )+35)\left(188 \cos (2 \theta )-19 \cos (4 \theta )+2455\right)\right).
  \end{split}
\end{equation}

\begin{equation}\label{LISAcross}
\begin{split}
  M^{ad}_{11} &=\frac{81\sqrt{3} \pi    f R \sin ^2(\alpha ) \sin ^2(2 \theta ) (7 \cos (\iota )+\cos (3 \iota ))}{256c }  \sin (2 \alpha -4 \phi ), \\
  M^{ad}_{12}    & =\frac{\sqrt{3} \pi f R \sin ^2(\alpha ) \sin (2 \theta ) (7 \cos (\iota )+\cos (3 \iota )) }{128c } \left(-27 \sin ^2(\theta ) \cos (2 \alpha -4 \phi )\right.\\
 &\left.+47 \cos (2 \theta )+161\right),\\
   M^{ad}_{14}    & = \frac{\sqrt{3} \pi    f R \sin ^2(\alpha ) \sin ^3(\iota )}{256c }  \left( -27 (10 \sin (2 \theta )+3 \sin (4 \theta )) \cos (4 \psi ) \sin (2 \alpha -4 \phi )        \right.\\
  &+2 \sin (\theta ) \cos ^2(\theta ) \sin (4 \psi ) \left(54 \cos (2 \alpha ) (\cos (2 \theta )+7) \cos (4 \phi )\right.\\
  &\left.\left.+54 \sin (2 \alpha ) (\cos (2 \theta )+7) \sin (4 \phi )-40 \sin ^2(\theta )\right)\right),         \\
   M^{ad}_{15}    & =  \frac{ \sqrt{3} \pi    f R \sin ^2(\alpha ) \sin (\theta ) \cos ^2(\theta ) (7 \cos (\iota )+\cos (3 \iota )) }{32c}\left(27 \sin ^2(\theta ) \cos (2 \alpha -4 \phi )\right.\\
  &\left.-5 \cos (2 \theta )+109\right),        \\
    M^{ad}_{16}    & =    \frac{\sqrt{3} \pi    f R \sin ^2(\alpha )}{2048c}   \left(108 (10 \sin (2 \theta )+3 \sin (4 \theta )) \sin ^4(\iota ) \sin (4 \psi ) \sin (2 \alpha -4 \phi )\right.\\
   &+4 \sin (\theta ) \cos ^2(\theta ) \left(2 \sin ^4(\iota ) \cos (4 \psi ) \left(54 \cos (2 \alpha ) (\cos (2 \theta )+7) \cos (4 \phi )\right.\right.\\
  &\left. +54 \sin (2 \alpha ) (\cos (2 \theta )+7) \sin (4 \phi )-40 \sin ^2(\theta )\right)\\
   &\left.\left.+(28 \cos (2 \iota )+\cos (4 \iota )+35) \left(-27 \sin ^2(\theta ) \cos (2 \alpha -4 \phi )+5 \cos (2 \theta )-109\right)\right)\right),    \\
   M^{ad}_{22}    & =  \frac{27\sqrt{3} \pi    f R \sin ^2(\alpha ) \sin ^4(\theta ) (7 \cos (\iota )+\cos (3 \iota ))}{64c}  \sin (2 \alpha -4 \phi ),        \\
   M^{ad}_{24}    & = \frac{\sqrt{3} \pi    f R \sin ^2(\alpha ) \sin ^2(\theta ) \sin ^3(\iota )}{128c}  \left(2 \cos (4 \psi ) \left(27 (3 \cos (2 \theta )+5) \cos (2 \alpha -4 \phi )\right)\right.\\
  &\left.                       +27 (15 \cos (\theta )+\cos (3 \theta )) \sin (4 \psi ) \sin (2 \alpha -4 \phi )       +       40 \cos (4 \psi ) \sin ^2(\theta )                           \right),         \\
   M^{ad}_{25}    & =\frac{27\sqrt{3} \pi    f R \sin ^2(\alpha ) \sin ^4(\theta ) \cos (\theta ) (7 \cos (\iota )+\cos (3 \iota ))}{32c}  \sin (2 \alpha -4 \phi ),          \\
   M^{ad}_{26}    & =\frac{ \sqrt{3} \pi    f R \sin ^2(\alpha ) \sin ^2(\theta )}{512c} \left(54 (15 \cos (\theta )+\cos (3 \theta )) \sin ^4(\iota ) \cos (4 \psi ) \sin (2 \alpha -4 \phi )\right.\\
  &-108\sin ^4(\iota ) \sin (4 \psi ) \cos (2 \alpha ) (3 \cos (2 \theta )+5) \cos (4 \phi )           \\
  &-27 \sin ^2(\theta ) \cos (\theta ) (28 \cos (2 \iota )+\cos (4 \iota )+35) \sin (2 \alpha -4 \phi )\\
  &\left.     -2 \sin ^4(\iota ) \sin (4 \psi ) \left(54 \sin (2 \alpha ) (3 \cos (2 \theta )+5) \sin (4 \phi )+40 \sin ^2(\theta )\right)      \right).          \\
\end{split}
\end{equation}
\end{widetext}

\section{The frequency dependence of the inverse of FIM} \label{InverseFIM}
\subsection{TianQin}
We can write the FIM as ${\bf \Gamma}=\textbf{H}+f^2\textbf{N}$, where $ \textbf{H}$ and $\textbf{N}$ are  independent of the frequency.
In the low frequency limit $f\rightarrow 0$,
\begin{equation}
\begin{split}
    \det {\bf \Gamma}&=\det(\textbf{H}+f^2\textbf{N})\\
    &=\det \textbf{H}+f^2~\text{Tr}~\left[ \text{adj}~(\textbf{H})~\textbf{N}\right]+O(f^4),
\end{split}
\end{equation}
where $\text{adj}~(\textbf{H})$ is the adjugate of the matrix $\textbf{H}$ and $\text{Tr}$ denotes the trace of the matrix,
so the inverse of the FIM is
\begin{equation}
\begin{split}
    {\bf \Gamma}^{-1}&=\frac{1}{\det {\bf \Gamma}} \text{adj}~({\bf \Gamma})\\
    &=\frac{\text{adj}~(\textbf{H}+f^2\textbf{N})}{\det \textbf{H}+f^2~\text{Tr}~\left[ \text{adj}~(\textbf{H})~\textbf{N}\right]}.
\end{split}
\end{equation}
If the matrix $\textbf{H}$ is non-singular then $\bf{\Gamma}^{-1}$ can be expanded as 
\begin{equation}
    {\bf \Gamma}^{-1}=\textbf{H}^{-1}+O(f^2).
\end{equation}
If the matrix $\textbf{H}$ is singular then $\bf{\Gamma}^{-1}$ can be expanded as
\begin{equation}
    {\bf \Gamma}^{-1}=\frac{\text{adj}~(\textbf{H})}{f^2~\text{Tr}~\left[ \text{adj}~(\textbf{H})~\textbf{N}\right]}+O(f^0).
\end{equation}

In order to discuss the high frequency limit, we explicitly write the matrices $\textbf{H}$ and $\textbf{N}$ as block matrices, then the FIM becomes
\begin{equation}
{\bf \Gamma}=\begin{pmatrix}
\textbf{Q}_{2\times2}, \textbf{W}_{2\times4}\\
\textbf{W}^T_{4\times2},\textbf{X}_{4\times4}
\end{pmatrix}+f^2
\begin{pmatrix}
\textbf{Y}_{2\times2}, \textbf{0}_{2\times4}\\
\textbf{0}_{4\times2},\textbf{0}_{4\times4}
\end{pmatrix},
\end{equation}
and the inverse of the FIM is
\begin{equation}
\begin{split}
     {\bf \Gamma}^{-1}&=\begin{pmatrix}
{\bf V}^{-1}&,& \ast\\
\ast&,&{\bf Z}^{-1}
\end{pmatrix},
\end{split}
\end{equation}
where
${\bf V}=
\textbf{Q}+f^2\textbf{Y}-\textbf{W}\textbf{X}^{-1}\textbf{W}^T$,
${\bf Z}=\textbf{X}-\textbf{W}^T(\textbf{Q}+f^2\textbf{Y})^{-1}\textbf{W}$,
the off diagonal elements are not specified because they are irrelevant for our discussion.
In the high frequency limit $f\rightarrow\infty$, we get
\begin{equation}
{\bf\Gamma}^{-1}=\begin{pmatrix}
f^{-2}\textbf{Y}^{-1}_{2\times2}&,& \ast\\
\ast&,&\textbf{X}^{-1}_{4\times4}
\end{pmatrix}.
\end{equation}

\subsection{LISA}
We can write the FIM as ${\bf \Gamma}=\textbf{H}+f{\bf K}+f^2\textbf{N}$, where $ \textbf{H}$, $ \textbf{K}$ and $\textbf{N}$ are  independent of the frequency.
In the low frequency limit $f\rightarrow 0$, the matrix $\textbf{H}$ is non-singular and $\bf{\Gamma}^{-1}$ can be expanded as 
\begin{equation}
    {\bf \Gamma}^{-1}=\textbf{H}^{-1}+O(f).
\end{equation}

In order to discuss the high frequency limit, we explicitly write the matrices $\textbf{H}$, $\textbf{K}$ and $\textbf{N}$ as block matrices, then the FIM becomes
\begin{equation}
{\bf \Gamma}=\begin{pmatrix}
\textbf{Q}&, &\textbf{W}\\
\textbf{W}^T&,&\textbf{X}
\end{pmatrix}+
\begin{pmatrix}
f{\bf B}+f^2\textbf{Y}&,& f{\bf C}\\
f{\bf C}^T&,&\textbf{0}
\end{pmatrix},
\end{equation}
and the inverse of the FIM is
\begin{equation}
\begin{split}
     {\bf \Gamma}^{-1}&=\begin{pmatrix}
{\bf V}^{-1}&,& \ast\\
\ast&,&{\bf Z}^{-1}
\end{pmatrix},
\end{split}
\end{equation}
where
${\bf V}=
\textbf{Q}+f\textbf{B}+f^2\textbf{Y}-(\textbf{W}+f\textbf{C})\textbf{X}^{-1}(\textbf{W}+f\textbf{C})^T$,
${\bf Z}=\textbf{X}-(\textbf{W}+f\textbf{C})^T(\textbf{Q}+f\textbf{B}+f^2\textbf{Y})^{-1}(\textbf{W}+f\textbf{C})$,
the off diagonal elements are not specified because they are irrelevant for our discussion.
In the high frequency limit $f\rightarrow\infty$, we get
\begin{equation*}
{\bf\Gamma}^{-1}=\begin{pmatrix}
f^{-2}(\textbf{Y}-\textbf{C}\textbf{X}^{-1}\textbf{C}^T)^{-1}&,& \ast\\
\ast&,&(\textbf{X}-\textbf{C}^T\textbf{Y}^{-1}\textbf{C})^{-1}
\end{pmatrix}.
\end{equation*}

\section{The FIM with four parameters} 
\label{TQ4paras}
\begin{figure}
  \centering
  \includegraphics[width=0.8\columnwidth]{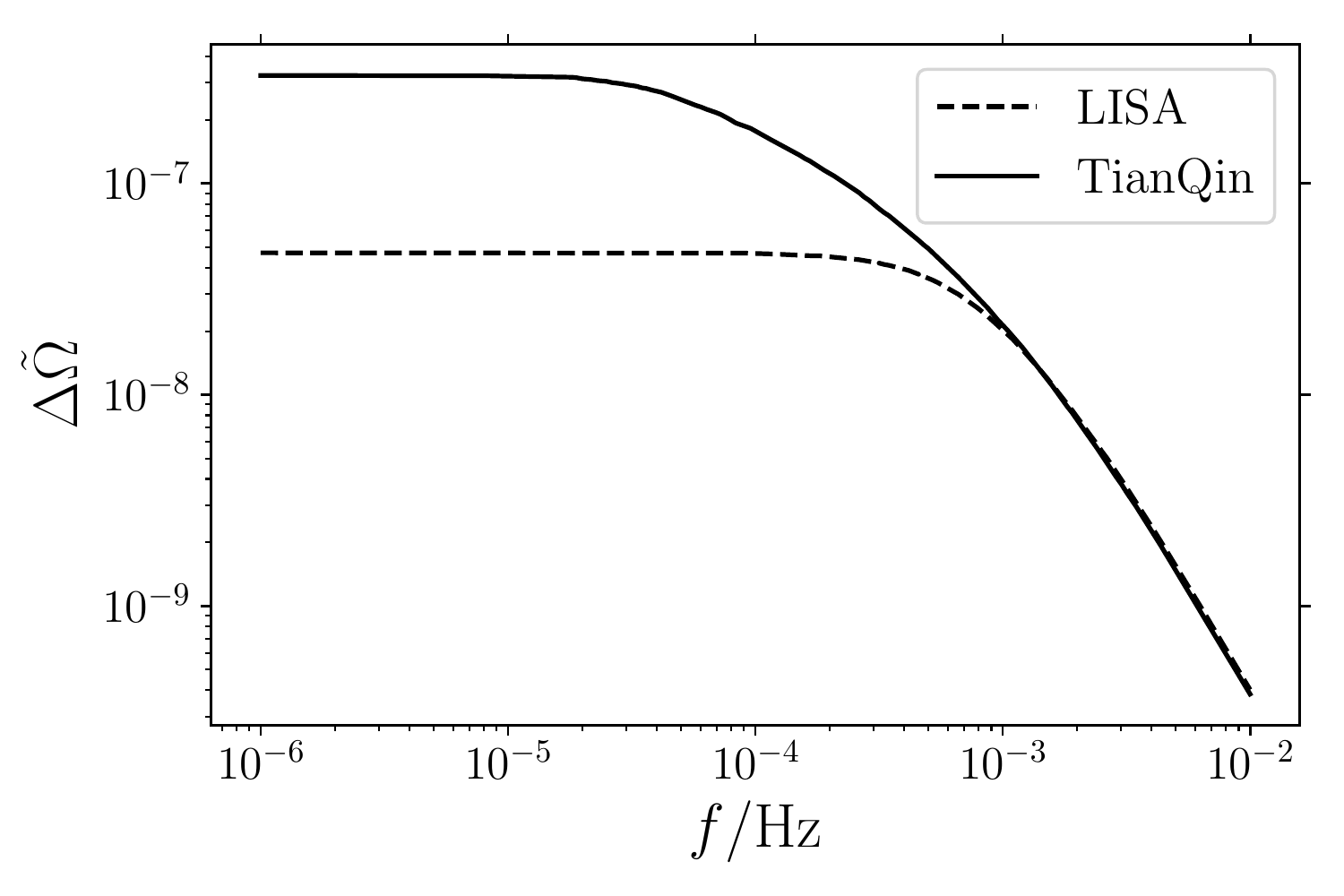}
  \caption{The medians of the normalized angular resolutions $\Delta \tilde\Omega$ 
  of LISA and TianQin
  as functions of frequency for monochromatic sources with four parameters.}\label{omegas4}
\end{figure}

\begin{figure*}
  \centering
  \includegraphics[width=0.9\textwidth]{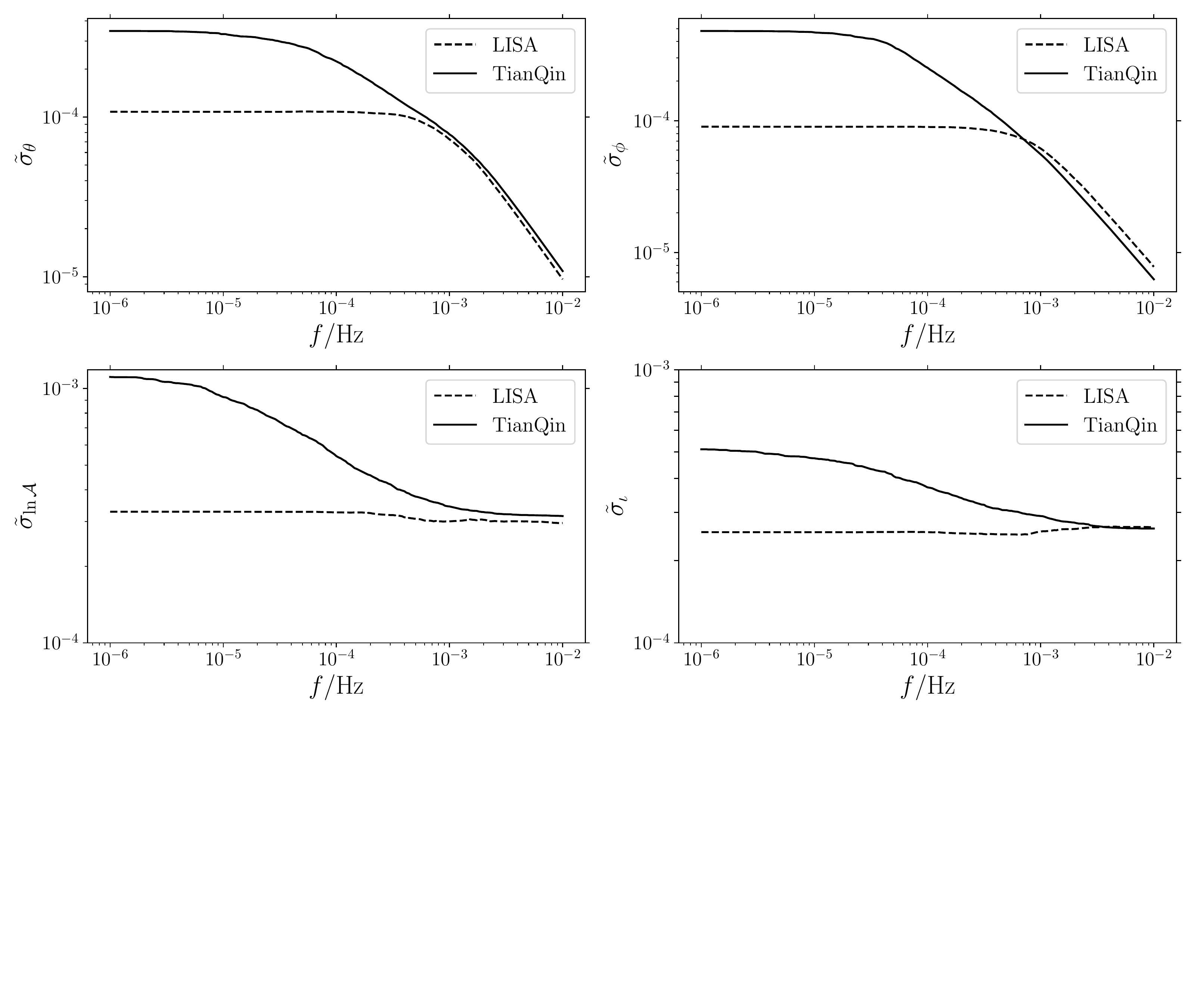}
  \caption{The medians of parameter estimation errors of LISA and TianQin as functions of frequency for monochromatic sources with four parameters in the low and medium frequency regimes.}
  \label{parametersLWaverage4}
\end{figure*}

\begin{figure*}
  \centering
  \includegraphics[width=0.9\textwidth]{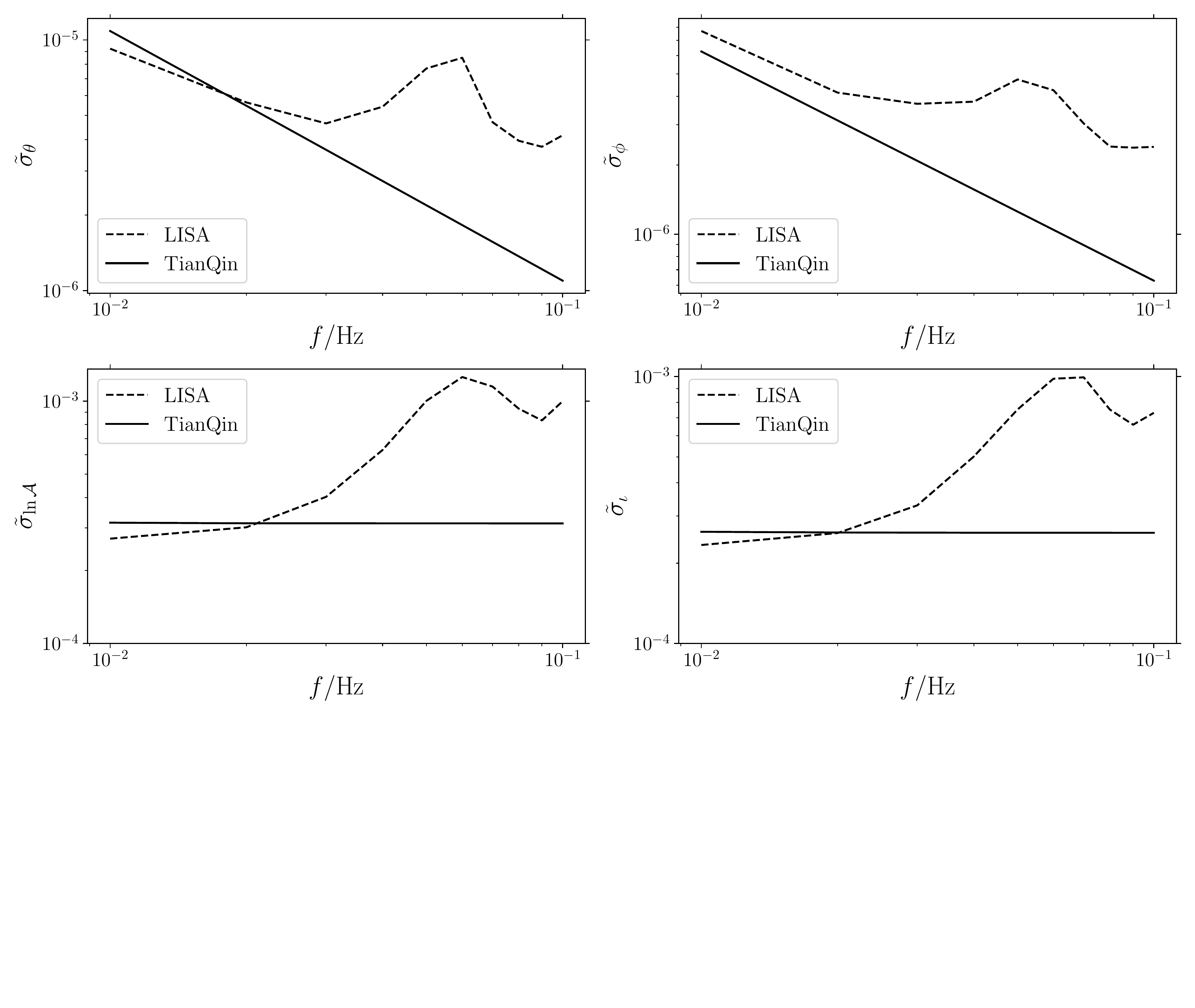}
  \caption{The medians of parameter estimation errors of LISA and TianQin as functions of frequency for monochromatic sources with four parameters in the high frequency regimes.}
  \label{hparameters4}
\end{figure*}

If we consider only four parameters of the monochromatic GW signal
\begin{equation}
    \boldsymbol{\xi}=\{ \theta, \phi,\ln{\mathcal{A}},\iota\},
\end{equation}
then the FIM in the LW approximation can be written in a compact form as
\begin{equation}
 \Gamma^{LW}=\Gamma^{am}+\Gamma^{dm},
\end{equation}
where
\begin{equation}
 \Gamma^{am}=\frac{\mathcal{A}^2T_{obs}  }{S_n(f)}F^{am},\quad
  \Gamma^{dm}=\frac{\mathcal{A}^2 T_{obs} }{S_n(f)}F^{dm},
\end{equation}
$F^{am}$ is a 4 by 4 submatrix of $M^{am}$ and independent of the frequency,
\begin{equation}
 F^{dm}=\left(\frac{2\pi f R}{c} \right)^2\begin{pmatrix}
\textbf{D}_{2\times2}, \textbf{0}_{2\times2}\\
\textbf{0}_{2\times2},\textbf{0}_{2\times2}
\end{pmatrix}.
\end{equation}
Let us chose $\psi_d=0$, we get
\begin{equation*}
|F^{am}|=\frac{36 \sin ^2(\theta)\sin ^2(\theta_d) \sin ^2(\iota ) \left[\cos (2 \theta_d)-\cos (2 \iota )\right]^2}{\sin ^2(\theta) \left(\sqrt{3} \sin (2 \phi)+\cos (2 \phi)\right)+\cos (2 \theta)+3}.
\end{equation*}
In general, $|F^{am}|\neq 0$, so the matrix $F^{am}$ is non-singular and independent of the frequency.
We find that in TianQin the total amplitude modulation $|M^{am}|=0$ for 
the 6 parameter set while  the total amplitude modulation $|F^{am}|\neq 0$ for 
the 4 parameter set.
Reducing two parameters ($\psi,\phi_0$) means that these two parameters can be determined as a prior by other ways before the GW detection.
The remaining four parameters can be detected by the rotational effect around the constant axis which points to the reference source.
But the six parameters are highly correlated and can not be discriminated just by the rotational effect around the constant axis for TianQin.
For LISA these six parameters can be discriminated 
by the rotation of the detector plane whose axis rotates around the normal vector of the ecliptic plane.
The medians of angular resolutions and
parameter estimation errors of LISA and TianQin in the low and medium frequency regimes are shown in Figs. \ref{omegas4} and \ref{parametersLWaverage4}, respectively.
The medians of parameter estimation errors of LISA and TianQin in the high frequency regime are shown  Fig. \ref{hparameters4}.
For LISA, the results with 4 parameters are similar to those with 6 parameters
except that the errors are smaller.
For TianQin, the results with 4 parameters are similar to those with 6 parameters
in the medium and high frequency regimes
except that the errors are smaller.
In the low frequency regime, the amplitude modulation due to the rotation of the spacecrafts around the fixed axis helps TianQin to reduce the parameter errors if we consider the four parameters only.

\section{The Bayesian analysis} 
\label{bayes1}
\begin{figure}
  \centering
  \includegraphics[width=0.98\columnwidth]{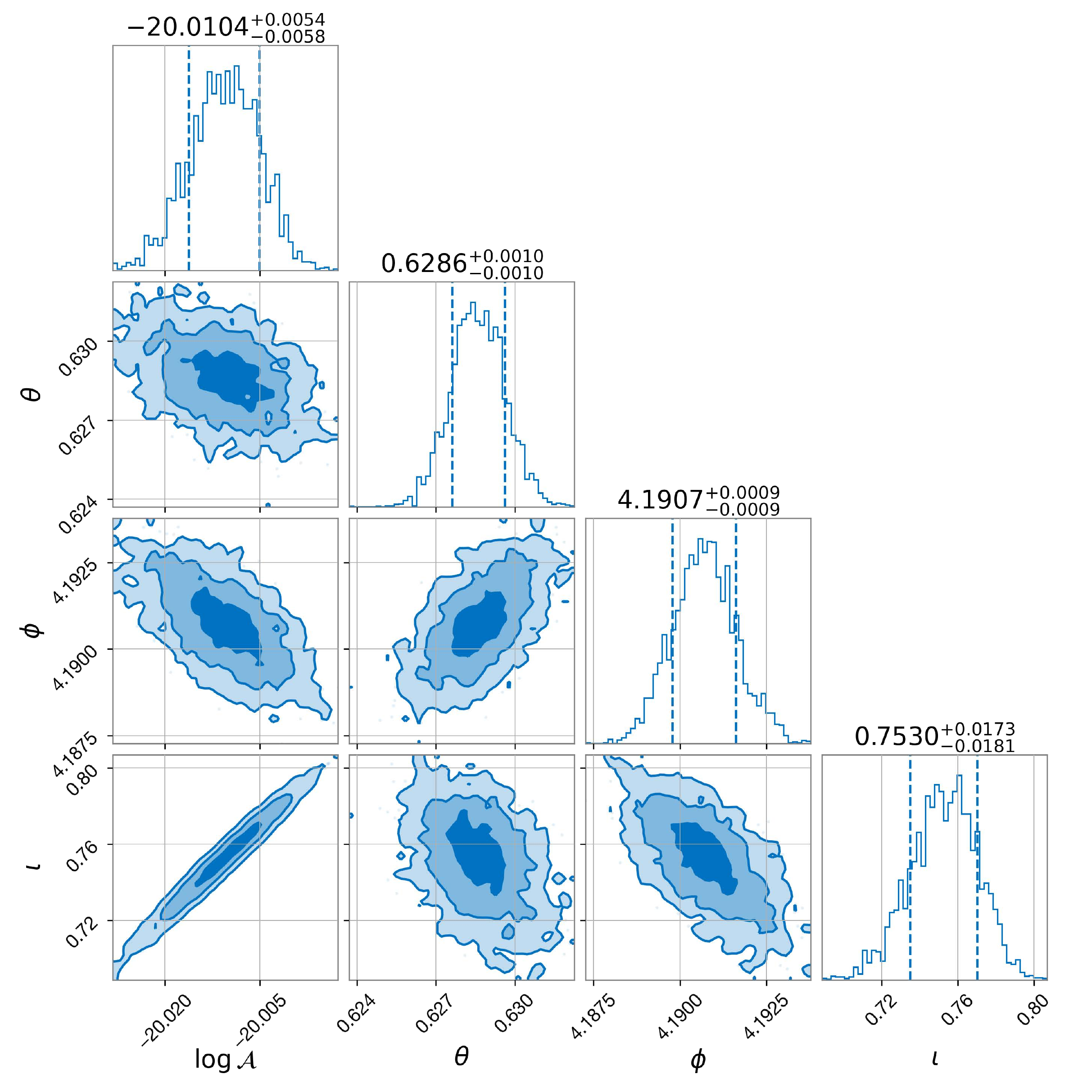}
  \caption{The parameter estimation errors using the Bayesian analysis for the simulated source with ($\mathcal{A}=10^{-20},\theta=\pi/5, \phi=4\pi/3, \iota=\pi/4, \psi=\pi/4, \phi_0=0, f=10^{-3}$).}
  \label{lisa_corner}
\end{figure}

The likelihood $\mathcal{L} = p\left(d|\bm{\xi}\right)$ for a given gravitational wave signal takes the form
\begin{equation}
\label{deflnL}
\ln{\mathcal{L}}=-\frac{1}{2}\left([H({\bm{\xi}})-d]|[H({\bm{\xi}})-d]\right),
\end{equation}
where the Hermitian inner product $(A|B)$ is defined in Eq. \eqref{gijproduct},
the data $d = H(\bm{\xi}_0)+n$ is the superposition of the gravitational wave signal $H(\bm{\xi}_0)$ for the true parameters $\bm{\xi}_0$ and the noise $n$ in the detector.
The posterior distribution for the parameters $\bm{\xi}$ is
\begin{equation}
\label{defBayes}
p\left( \bm{\xi} | d \right) = \frac{p(d | \bm{\xi})p(\bm{\xi})}{p(d)},
\end{equation}
where $p(\bm{\xi})$ is the prior on the parameters and $p(d)$ is the evidence.
In our analysis, we simulate signals in LISA for the source with parameters ($\mathcal{A}=10^{-20},\theta=\pi/5, \phi=4\pi/3, \iota=\pi/4, \psi=\pi/4, \phi_0=0, f=10^{-3}$)  and perform Bayesian analyses with Eq.~\eqref{deflnL} and ~\eqref{defBayes} to obtain the posteriors of the physical parameter $\bm{\xi}$.
The signal to noise ratio for the source is SNR=955.
The results on the parameter estimations are shown in Fig. \ref{lisa_corner} and the comparison of the Bayesian analysis with the FIM is shown in Table \ref{FB}.
The results show that the parameter estimation errors with he Bayesian analysis are $3-6$ times larger that those with the FIM.

\begin{table}[htp]
\centering
\resizebox{0.95\columnwidth}{!}{
 	\begin{tabular}{|c|c|c|c|c|}
\hline
	Method & $\sigma_{\theta}(10^{-3})$ & $\sigma_{\phi}(10^{-3})$ & $\sigma_{\log\mathcal{A}}(10^{-3})$& $\sigma_{\iota}(10^{-3})$ \\ \hline
	Bayesian & 1.0  & 0.9  &5.5  &18  \\  \hline
	FIM      & 0.30  &  0.35 &  1.3 & 3.4  \\ \hline 
\end{tabular}
}
\caption{The parameter estimation errors using the FIM and the Bayesian analysis  for the fiducial source with ($\mathcal{A}=10^{-20},\theta=\pi/5, \phi=4\pi/3, \iota=\pi/4, \psi=\pi/4, \phi_0=0, f=10^{-3}$).}
\label{FB}
\end{table}


\begin{thebibliography}{50}%
\makeatletter
\providecommand \@ifxundefined [1]{%
 \@ifx{#1\undefined}
}%
\providecommand \@ifnum [1]{%
 \ifnum #1\expandafter \@firstoftwo
 \else \expandafter \@secondoftwo
 \fi
}%
\providecommand \@ifx [1]{%
 \ifx #1\expandafter \@firstoftwo
 \else \expandafter \@secondoftwo
 \fi
}%
\providecommand \natexlab [1]{#1}%
\providecommand \enquote  [1]{``#1''}%
\providecommand \bibnamefont  [1]{#1}%
\providecommand \bibfnamefont [1]{#1}%
\providecommand \citenamefont [1]{#1}%
\providecommand \href@noop [0]{\@secondoftwo}%
\providecommand \href [0]{\begingroup \@sanitize@url \@href}%
\providecommand \@href[1]{\@@startlink{#1}\@@href}%
\providecommand \@@href[1]{\endgroup#1\@@endlink}%
\providecommand \@sanitize@url [0]{\catcode `\\12\catcode `\$12\catcode
  `\&12\catcode `\#12\catcode `\^12\catcode `\_12\catcode `\%12\relax}%
\providecommand \@@startlink[1]{}%
\providecommand \@@endlink[0]{}%
\providecommand \url  [0]{\begingroup\@sanitize@url \@url }%
\providecommand \@url [1]{\endgroup\@href {#1}{\urlprefix }}%
\providecommand \urlprefix  [0]{URL }%
\providecommand \Eprint [0]{\href }%
\providecommand \doibase [0]{https://doi.org/}%
\providecommand \selectlanguage [0]{\@gobble}%
\providecommand \bibinfo  [0]{\@secondoftwo}%
\providecommand \bibfield  [0]{\@secondoftwo}%
\providecommand \translation [1]{[#1]}%
\providecommand \BibitemOpen [0]{}%
\providecommand \bibitemStop [0]{}%
\providecommand \bibitemNoStop [0]{.\EOS\space}%
\providecommand \EOS [0]{\spacefactor3000\relax}%
\providecommand \BibitemShut  [1]{\csname bibitem#1\endcsname}%
\let\auto@bib@innerbib\@empty
\bibitem [{\citenamefont {Abbott}\ {\it
  et~al.}(2016{\natexlab{a}})\citenamefont {Abbott} {\it
  et~al.}}]{TheLIGOScientific:2016agk}%
  \BibitemOpen
  \bibfield  {author} {\bibinfo {author} {\bibfnamefont {B.~P.}\ \bibnamefont
  {Abbott}} {\it et~al.} (\bibinfo {collaboration} {LIGO Scientific, Virgo}),\
  }\bibinfo {title} {{GW150914: The Advanced LIGO Detectors in the Era of First
  Discoveries}},\ \href {https://doi.org/10.1103/PhysRevLett.116.131103}
  {\bibfield  {journal} {\bibinfo  {journal} {Phys. Rev. Lett.}\ }\textbf
  {\bibinfo {volume} {116}},\ \bibinfo {pages} {131103} (\bibinfo {year}
  {2016}{\natexlab{a}})}\BibitemShut {NoStop}%
\bibitem [{\citenamefont {Abbott}\ {\it
  et~al.}(2016{\natexlab{b}})\citenamefont {Abbott} {\it
  et~al.}}]{Abbott:2016blz}%
  \BibitemOpen
  \bibfield  {author} {\bibinfo {author} {\bibfnamefont {B.~P.}\ \bibnamefont
  {Abbott}} {\it et~al.} (\bibinfo {collaboration} {LIGO Scientific, Virgo}),\
  }\bibinfo {title} {{Observation of Gravitational Waves from a Binary Black
  Hole Merger}},\ \href {https://doi.org/10.1103/PhysRevLett.116.061102}
  {\bibfield  {journal} {\bibinfo  {journal} {Phys. Rev. Lett.}\ }\textbf
  {\bibinfo {volume} {116}},\ \bibinfo {pages} {061102} (\bibinfo {year}
  {2016}{\natexlab{b}})}\BibitemShut {NoStop}%
\bibitem [{\citenamefont {Abbott}\ {\it
  et~al.}(2016{\natexlab{c}})\citenamefont {Abbott} {\it
  et~al.}}]{Abbott:2016nmj}%
  \BibitemOpen
  \bibfield  {author} {\bibinfo {author} {\bibfnamefont {B.~P.}\ \bibnamefont
  {Abbott}} {\it et~al.} (\bibinfo {collaboration} {LIGO Scientific, Virgo}),\
  }\bibinfo {title} {{GW151226: Observation of Gravitational Waves from a
  22-Solar-Mass Binary Black Hole Coalescence}},\ \href
  {https://doi.org/10.1103/PhysRevLett.116.241103} {\bibfield  {journal}
  {\bibinfo  {journal} {Phys. Rev. Lett.}\ }\textbf {\bibinfo {volume} {116}},\
  \bibinfo {pages} {241103} (\bibinfo {year} {2016}{\natexlab{c}})}\BibitemShut
  {NoStop}%
\bibitem [{\citenamefont {Abbott}\ {\it
  et~al.}(2017{\natexlab{a}})\citenamefont {Abbott} {\it
  et~al.}}]{Abbott:2017vtc}%
  \BibitemOpen
  \bibfield  {author} {\bibinfo {author} {\bibfnamefont {B.~P.}\ \bibnamefont
  {Abbott}} {\it et~al.} (\bibinfo {collaboration} {LIGO Scientific, VIRGO}),\
  }\bibinfo {title} {{GW170104: Observation of a 50-Solar-Mass Binary Black
  Hole Coalescence at Redshift 0.2}},\ \href
  {https://doi.org/10.1103/PhysRevLett.118.221101} {\bibfield  {journal}
  {\bibinfo  {journal} {Phys. Rev. Lett.}\ }\textbf {\bibinfo {volume} {118}},\
  \bibinfo {pages} {221101} (\bibinfo {year} {2017}{\natexlab{a}})},\ \bibinfo
  {note} {[Erratum: Phys.Rev.Lett. 121, 129901 (2018)]}\BibitemShut {NoStop}%
\bibitem [{\citenamefont {Abbott}\ {\it
  et~al.}(2017{\natexlab{b}})\citenamefont {Abbott} {\it
  et~al.}}]{Abbott:2017oio}%
  \BibitemOpen
  \bibfield  {author} {\bibinfo {author} {\bibfnamefont {B.~P.}\ \bibnamefont
  {Abbott}} {\it et~al.} (\bibinfo {collaboration} {LIGO Scientific, Virgo}),\
  }\bibinfo {title} {{GW170814: A Three-Detector Observation of Gravitational
  Waves from a Binary Black Hole Coalescence}},\ \href
  {https://doi.org/10.1103/PhysRevLett.119.141101} {\bibfield  {journal}
  {\bibinfo  {journal} {Phys. Rev. Lett.}\ }\textbf {\bibinfo {volume} {119}},\
  \bibinfo {pages} {141101} (\bibinfo {year} {2017}{\natexlab{b}})}\BibitemShut
  {NoStop}%
\bibitem [{\citenamefont {Abbott}\ {\it
  et~al.}(2017{\natexlab{c}})\citenamefont {Abbott} {\it
  et~al.}}]{TheLIGOScientific:2017qsa}%
  \BibitemOpen
  \bibfield  {author} {\bibinfo {author} {\bibfnamefont {B.~P.}\ \bibnamefont
  {Abbott}} {\it et~al.} (\bibinfo {collaboration} {LIGO Scientific, Virgo}),\
  }\bibinfo {title} {{GW170817: Observation of Gravitational Waves from a
  Binary Neutron Star Inspiral}},\ \href
  {https://doi.org/10.1103/PhysRevLett.119.161101} {\bibfield  {journal}
  {\bibinfo  {journal} {Phys. Rev. Lett.}\ }\textbf {\bibinfo {volume} {119}},\
  \bibinfo {pages} {161101} (\bibinfo {year} {2017}{\natexlab{c}})}\BibitemShut
  {NoStop}%
\bibitem [{\citenamefont {Abbott}\ {\it
  et~al.}(2017{\natexlab{d}})\citenamefont {Abbott} {\it
  et~al.}}]{Abbott:2017gyy}%
  \BibitemOpen
  \bibfield  {author} {\bibinfo {author} {\bibfnamefont {B.~. P.~.}\
  \bibnamefont {Abbott}} {\it et~al.} (\bibinfo {collaboration} {LIGO
  Scientific, Virgo}),\ }\bibinfo {title} {{GW170608: Observation of a
  19-solar-mass Binary Black Hole Coalescence}},\ \href
  {https://doi.org/10.3847/2041-8213/aa9f0c} {\bibfield  {journal} {\bibinfo
  {journal} {Astrophys. J. Lett.}\ }\textbf {\bibinfo {volume} {851}},\
  \bibinfo {pages} {L35} (\bibinfo {year} {2017}{\natexlab{d}})}\BibitemShut
  {NoStop}%
\bibitem [{\citenamefont {Abbott}\ {\it et~al.}(2019)\citenamefont {Abbott}
  {\it et~al.}}]{LIGOScientific:2018mvr}%
  \BibitemOpen
  \bibfield  {author} {\bibinfo {author} {\bibfnamefont {B.~P.}\ \bibnamefont
  {Abbott}} {\it et~al.} (\bibinfo {collaboration} {LIGO Scientific, Virgo}),\
  }\bibinfo {title} {{GWTC-1: A Gravitational-Wave Transient Catalog of Compact
  Binary Mergers Observed by LIGO and Virgo during the First and Second
  Observing Runs}},\ \href {https://doi.org/10.1103/PhysRevX.9.031040}
  {\bibfield  {journal} {\bibinfo  {journal} {Phys. Rev. X}\ }\textbf {\bibinfo
  {volume} {9}},\ \bibinfo {pages} {031040} (\bibinfo {year}
  {2019})}\BibitemShut {NoStop}%
\bibitem [{\citenamefont {Abbott}\ {\it
  et~al.}(2020{\natexlab{a}})\citenamefont {Abbott} {\it
  et~al.}}]{Abbott:2020uma}%
  \BibitemOpen
  \bibfield  {author} {\bibinfo {author} {\bibfnamefont {B.~P.}\ \bibnamefont
  {Abbott}} {\it et~al.} (\bibinfo {collaboration} {LIGO Scientific, Virgo}),\
  }\bibinfo {title} {{GW190425: Observation of a Compact Binary Coalescence
  with Total Mass $\sim 3.4 M_{\odot}$}},\ \href
  {https://doi.org/10.3847/2041-8213/ab75f5} {\bibfield  {journal} {\bibinfo
  {journal} {Astrophys. J. Lett.}\ }\textbf {\bibinfo {volume} {892}},\
  \bibinfo {pages} {L3} (\bibinfo {year} {2020}{\natexlab{a}})}\BibitemShut
  {NoStop}%
\bibitem [{\citenamefont {Abbott}\ {\it
  et~al.}(2020{\natexlab{b}})\citenamefont {Abbott} {\it
  et~al.}}]{LIGOScientific:2020stg}%
  \BibitemOpen
  \bibfield  {author} {\bibinfo {author} {\bibfnamefont {R.}~\bibnamefont
  {Abbott}} {\it et~al.} (\bibinfo {collaboration} {LIGO Scientific, Virgo}),\
  }\bibinfo {title} {{GW190412: Observation of a Binary-Black-Hole Coalescence
  with Asymmetric Masses}},\ \href
  {https://doi.org/10.1103/PhysRevD.102.043015} {\bibfield  {journal} {\bibinfo
   {journal} {Phys. Rev. D}\ }\textbf {\bibinfo {volume} {102}},\ \bibinfo
  {pages} {043015} (\bibinfo {year} {2020}{\natexlab{b}})}\BibitemShut
  {NoStop}%
\bibitem [{\citenamefont {Abbott}\ {\it
  et~al.}(2020{\natexlab{c}})\citenamefont {Abbott} {\it
  et~al.}}]{Abbott:2020khf}%
  \BibitemOpen
  \bibfield  {author} {\bibinfo {author} {\bibfnamefont {R.}~\bibnamefont
  {Abbott}} {\it et~al.} (\bibinfo {collaboration} {LIGO Scientific, Virgo}),\
  }\bibinfo {title} {{GW190814: Gravitational Waves from the Coalescence of a
  23 Solar Mass Black Hole with a 2.6 Solar Mass Compact Object}},\ \href
  {https://doi.org/10.3847/2041-8213/ab960f} {\bibfield  {journal} {\bibinfo
  {journal} {Astrophys. J. Lett.}\ }\textbf {\bibinfo {volume} {896}},\
  \bibinfo {pages} {L44} (\bibinfo {year} {2020}{\natexlab{c}})}\BibitemShut
  {NoStop}%
\bibitem [{\citenamefont {Abbott}\ {\it
  et~al.}(2020{\natexlab{d}})\citenamefont {Abbott} {\it et~al.}}]{Abbott:2020tfl}%
  \BibitemOpen
  \bibfield  {author} {\bibinfo {author} {\bibfnamefont {R.}~\bibnamefont
  {Abbott}} {\it et~al.} (\bibinfo {collaboration} {LIGO Scientific, Virgo}),\
  }\bibinfo {title} {{GW190521: A Binary Black Hole Merger with a Total Mass of
  $150 M_{\odot}$}},\ \href {https://doi.org/10.1103/PhysRevLett.125.101102}
  {\bibfield  {journal} {\bibinfo  {journal} {Phys. Rev. Lett.}\ }\textbf
  {\bibinfo {volume} {125}},\ \bibinfo {pages} {101102} (\bibinfo {year}
  {2020}{\natexlab{d}})}\BibitemShut {NoStop}%
\bibitem [{\citenamefont {Abbott}\ {\it
  et~al.}(2020{\natexlab{e}})\citenamefont {Abbott} {\it
  et~al.}}]{Abbott:2020niy}%
  \BibitemOpen
  \bibfield  {author} {\bibinfo {author} {\bibfnamefont {R.}~\bibnamefont
  {Abbott}} {\it et~al.} (\bibinfo {collaboration} {LIGO Scientific, Virgo}),\
  }\bibinfo {title} {{GWTC-2: Compact Binary Coalescences Observed by LIGO and
  Virgo During the First Half of the Third Observing Run}},\ \Eprint
  {https://arxiv.org/abs/2010.14527} {arXiv:2010.14527 [gr-qc]} \BibitemShut
  {NoStop}%
\bibitem [{\citenamefont {Danzmann}(1997)}]{Danzmann:1997hm}%
  \BibitemOpen
  \bibfield  {author} {\bibinfo {author} {\bibfnamefont {K.}~\bibnamefont
  {Danzmann}},\ }\bibinfo {title} {{LISA: An ESA cornerstone mission for a
  gravitational wave observatory}},\ \href
  {https://doi.org/10.1088/0264-9381/14/6/002} {\bibfield  {journal} {\bibinfo
  {journal} {Class. Quant. Grav.}\ }\textbf {\bibinfo {volume} {14}},\ \bibinfo
  {pages} {1399} (\bibinfo {year} {1997})}\BibitemShut {NoStop}%
\bibitem [{\citenamefont {Amaro-Seoane}\ {\it et~al.}(2017)\citenamefont
  {Amaro-Seoane} {\it et~al.}}]{Audley:2017drz}%
  \BibitemOpen
  \bibfield  {author} {\bibinfo {author} {\bibfnamefont {P.}~\bibnamefont
  {Amaro-Seoane}} {\it et~al.} (\bibinfo {collaboration} {LISA}),\ }\bibinfo
  {title} {{Laser Interferometer Space Antenna}},\ \Eprint
  {https://arxiv.org/abs/1702.00786} {arXiv:1702.00786 [astro-ph.IM]}
  \BibitemShut {NoStop}%
\bibitem [{\citenamefont {Hu}\ and\ \citenamefont {Wu}(2017)}]{Hu:2017mde}%
  \BibitemOpen
  \bibfield  {author} {\bibinfo {author} {\bibfnamefont {W.-R.}\ \bibnamefont
  {Hu}}\ and\ \bibinfo {author} {\bibfnamefont {Y.-L.}\ \bibnamefont {Wu}},\
  }\bibinfo {title} {{The Taiji Program in Space for gravitational wave physics
  and the nature of gravity}},\ \href {https://doi.org/10.1093/nsr/nwx116}
  {\bibfield  {journal} {\bibinfo  {journal} {Natl. Sci. Rev.}\ }\textbf
  {\bibinfo {volume} {4}},\ \bibinfo {pages} {685} (\bibinfo {year}
  {2017})}\BibitemShut {NoStop}%
\bibitem [{\citenamefont {Luo}\ {\it et~al.}(2016)\citenamefont {Luo} {\it
  et~al.}}]{Luo:2015ght}%
  \BibitemOpen
  \bibfield  {author} {\bibinfo {author} {\bibfnamefont {J.}~\bibnamefont
  {Luo}} {\it et~al.} (\bibinfo {collaboration} {TianQin}),\ }\bibinfo {title}
  {{TianQin: a space-borne gravitational wave detector}},\ \href
  {https://doi.org/10.1088/0264-9381/33/3/035010} {\bibfield  {journal}
  {\bibinfo  {journal} {Class. Quant. Grav.}\ }\textbf {\bibinfo {volume}
  {33}},\ \bibinfo {pages} {035010} (\bibinfo {year} {2016})}\BibitemShut
  {NoStop}%
\bibitem [{\citenamefont {Schutz}(1986)}]{Schutz:1986gp}%
  \BibitemOpen
  \bibfield  {author} {\bibinfo {author} {\bibfnamefont {B.~F.}\ \bibnamefont
  {Schutz}},\ }\bibinfo {title} {{Determining the Hubble Constant from
  Gravitational Wave Observations}},\ \href {https://doi.org/10.1038/323310a0}
  {\bibfield  {journal} {\bibinfo  {journal} {Nature}\ }\textbf {\bibinfo
  {volume} {323}},\ \bibinfo {pages} {310} (\bibinfo {year}
  {1986})}\BibitemShut {NoStop}%
\bibitem [{\citenamefont {Holz}\ and\ \citenamefont
  {Hughes}(2005)}]{Holz:2005df}%
  \BibitemOpen
  \bibfield  {author} {\bibinfo {author} {\bibfnamefont {D.~E.}\ \bibnamefont
  {Holz}}\ and\ \bibinfo {author} {\bibfnamefont {S.~A.}\ \bibnamefont
  {Hughes}},\ }\bibinfo {title} {{Using gravitational-wave standard sirens}},\
  \href {https://doi.org/10.1086/431341} {\bibfield  {journal} {\bibinfo
  {journal} {Astrophys. J.}\ }\textbf {\bibinfo {volume} {629}},\ \bibinfo
  {pages} {15} (\bibinfo {year} {2005})}\BibitemShut {NoStop}%
\bibitem [{\citenamefont {Riess}\ {\it et~al.}(2019)\citenamefont {Riess},
  \citenamefont {Casertano}, \citenamefont {Yuan}, \citenamefont {Macri},\ and\
  \citenamefont {Scolnic}}]{Riess:2019cxk}%
  \BibitemOpen
  \bibfield  {author} {\bibinfo {author} {\bibfnamefont {A.~G.}\ \bibnamefont
  {Riess}}, \bibinfo {author} {\bibfnamefont {S.}~\bibnamefont {Casertano}},
  \bibinfo {author} {\bibfnamefont {W.}~\bibnamefont {Yuan}}, \bibinfo {author}
  {\bibfnamefont {L.~M.}\ \bibnamefont {Macri}},\ and\ \bibinfo {author}
  {\bibfnamefont {D.}~\bibnamefont {Scolnic}},\ }\bibinfo {title} {{Large
  Magellanic Cloud Cepheid Standards Provide a 1\% Foundation for the
  Determination of the Hubble Constant and Stronger Evidence for Physics beyond
  $\Lambda$CDM}},\ \href {https://doi.org/10.3847/1538-4357/ab1422} {\bibfield
  {journal} {\bibinfo  {journal} {Astrophys. J.}\ }\textbf {\bibinfo {volume}
  {876}},\ \bibinfo {pages} {85} (\bibinfo {year} {2019})}\BibitemShut
  {NoStop}%
\bibitem [{\citenamefont {Cutler}\ and\ \citenamefont
  {Flanagan}(1994)}]{Cutler:1994ys}%
  \BibitemOpen
  \bibfield  {author} {\bibinfo {author} {\bibfnamefont {C.}~\bibnamefont
  {Cutler}}\ and\ \bibinfo {author} {\bibfnamefont {E.~E.}\ \bibnamefont
  {Flanagan}},\ }\bibinfo {title} {{Gravitational waves from merging compact
  binaries: How accurately can one extract the binary's parameters from the
  inspiral wave form?}},\ \href {https://doi.org/10.1103/PhysRevD.49.2658}
  {\bibfield  {journal} {\bibinfo  {journal} {Phys. Rev. D}\ }\textbf {\bibinfo
  {volume} {49}},\ \bibinfo {pages} {2658} (\bibinfo {year}
  {1994})}\BibitemShut {NoStop}%
\bibitem [{\citenamefont {Peterseim}\ {\it et~al.}(1996)\citenamefont
  {Peterseim}, \citenamefont {Jennrich},\ and\ \citenamefont
  {Danzmann}}]{Peterseim:1996cw}%
  \BibitemOpen
  \bibfield  {author} {\bibinfo {author} {\bibfnamefont {M.}~\bibnamefont
  {Peterseim}}, \bibinfo {author} {\bibfnamefont {O.}~\bibnamefont
  {Jennrich}},\ and\ \bibinfo {author} {\bibfnamefont {K.}~\bibnamefont
  {Danzmann}},\ }\bibinfo {title} {{Accuracy of parameter estimation of
  gravitational waves with LISA}},\ \href
  {https://doi.org/10.1088/0264-9381/13/11A/037} {\bibfield  {journal}
  {\bibinfo  {journal} {Class. Quant. Grav.}\ }\textbf {\bibinfo {volume}
  {13}},\ \bibinfo {pages} {A279} (\bibinfo {year} {1996})}\BibitemShut
  {NoStop}%
\bibitem [{\citenamefont {Peterseim}\ {\it et~al.}(1997)\citenamefont
  {Peterseim}, \citenamefont {Jennrich}, \citenamefont {Danzmann},\ and\
  \citenamefont {Schutz}}]{Peterseim:1997ic}%
  \BibitemOpen
  \bibfield  {author} {\bibinfo {author} {\bibfnamefont {M.}~\bibnamefont
  {Peterseim}}, \bibinfo {author} {\bibfnamefont {O.}~\bibnamefont {Jennrich}},
  \bibinfo {author} {\bibfnamefont {K.}~\bibnamefont {Danzmann}},\ and\
  \bibinfo {author} {\bibfnamefont {B.~F.}\ \bibnamefont {Schutz}},\ }\bibinfo
  {title} {{Angular resolution of LISA}},\ \href
  {https://doi.org/10.1088/0264-9381/14/6/019} {\bibfield  {journal} {\bibinfo
  {journal} {Class. Quant. Grav.}\ }\textbf {\bibinfo {volume} {14}},\ \bibinfo
  {pages} {1507} (\bibinfo {year} {1997})}\BibitemShut {NoStop}%
\bibitem [{\citenamefont {Cutler}(1998)}]{Cutler:1997ta}%
  \BibitemOpen
  \bibfield  {author} {\bibinfo {author} {\bibfnamefont {C.}~\bibnamefont
  {Cutler}},\ }\bibinfo {title} {{Angular resolution of the LISA gravitational
  wave detector}},\ \href {https://doi.org/10.1103/PhysRevD.57.7089} {\bibfield
   {journal} {\bibinfo  {journal} {Phys. Rev. D}\ }\textbf {\bibinfo {volume}
  {57}},\ \bibinfo {pages} {7089} (\bibinfo {year} {1998})}\BibitemShut
  {NoStop}%
\bibitem [{\citenamefont {Cutler}\ and\ \citenamefont
  {Vecchio}(1998)}]{Cutler:1998muh}%
  \BibitemOpen
  \bibfield  {author} {\bibinfo {author} {\bibfnamefont {C.}~\bibnamefont
  {Cutler}}\ and\ \bibinfo {author} {\bibfnamefont {A.}~\bibnamefont
  {Vecchio}},\ }\bibinfo {title} {{LISA\textquoteright{}s angular resolution
  for monochromatic sources}},\ \href {https://doi.org/10.1063/1.57427}
  {\bibfield  {journal} {\bibinfo  {journal} {AIP Conf. Proc.}\ }\textbf
  {\bibinfo {volume} {456}},\ \bibinfo {pages} {95} (\bibinfo {year}
  {1998})}\BibitemShut {NoStop}%
\bibitem [{\citenamefont {Moore}\ and\ \citenamefont
  {Hellings}(2000)}]{Moore:1999zw}%
  \BibitemOpen
  \bibfield  {author} {\bibinfo {author} {\bibfnamefont {T.~A.}\ \bibnamefont
  {Moore}}\ and\ \bibinfo {author} {\bibfnamefont {R.~W.}\ \bibnamefont
  {Hellings}},\ }\bibinfo {title} {{The Angular resolution of space based
  gravitational wave detectors}},\ \href
  {https://doi.org/10.1103/PhysRevD.65.062001} {\bibfield  {journal} {\bibinfo
  {journal} {AIP Conf. Proc.}\ }\textbf {\bibinfo {volume} {523}},\ \bibinfo
  {pages} {255} (\bibinfo {year} {2000})}\BibitemShut {NoStop}%
\bibitem [{\citenamefont {Barack}\ and\ \citenamefont
  {Cutler}(2004)}]{Barack:2003fp}%
  \BibitemOpen
  \bibfield  {author} {\bibinfo {author} {\bibfnamefont {L.}~\bibnamefont
  {Barack}}\ and\ \bibinfo {author} {\bibfnamefont {C.}~\bibnamefont
  {Cutler}},\ }\bibinfo {title} {{LISA capture sources: Approximate waveforms,
  signal-to-noise ratios, and parameter estimation accuracy}},\ \href
  {https://doi.org/10.1103/PhysRevD.69.082005} {\bibfield  {journal} {\bibinfo
  {journal} {Phys. Rev. D}\ }\textbf {\bibinfo {volume} {69}},\ \bibinfo
  {pages} {082005} (\bibinfo {year} {2004})}\BibitemShut {NoStop}%
\bibitem [{\citenamefont {Porter}\ and\ \citenamefont
  {Cornish}(2008)}]{Porter:2008kn}%
  \BibitemOpen
  \bibfield  {author} {\bibinfo {author} {\bibfnamefont {E.~K.}\ \bibnamefont
  {Porter}}\ and\ \bibinfo {author} {\bibfnamefont {N.~J.}\ \bibnamefont
  {Cornish}},\ }\bibinfo {title} {{The Effect of Higher Harmonic Corrections on
  the Detection of massive black hole binaries with LISA}},\ \href
  {https://doi.org/10.1103/PhysRevD.78.064005} {\bibfield  {journal} {\bibinfo
  {journal} {Phys. Rev. D}\ }\textbf {\bibinfo {volume} {78}},\ \bibinfo
  {pages} {064005} (\bibinfo {year} {2008})}\BibitemShut {NoStop}%
\bibitem [{\citenamefont {Blaut}(2011)}]{Blaut:2011zz}%
  \BibitemOpen
  \bibfield  {author} {\bibinfo {author} {\bibfnamefont {A.}~\bibnamefont
  {Blaut}},\ }\bibinfo {title} {{Accuracy of estimation of parameters with
  LISA}},\ \href {https://doi.org/10.1103/PhysRevD.83.083006} {\bibfield
  {journal} {\bibinfo  {journal} {Phys. Rev. D}\ }\textbf {\bibinfo {volume}
  {83}},\ \bibinfo {pages} {083006} (\bibinfo {year} {2011})}\BibitemShut
  {NoStop}%
\bibitem [{\citenamefont {Ruan}\ {\it et~al.}(2019)\citenamefont {Ruan},
  \citenamefont {Liu}, \citenamefont {Guo}, \citenamefont {Wu},\ and\
  \citenamefont {Cai}}]{Ruan:2019tje}%
  \BibitemOpen
  \bibfield  {author} {\bibinfo {author} {\bibfnamefont {W.-H.}\ \bibnamefont
  {Ruan}}, \bibinfo {author} {\bibfnamefont {C.}~\bibnamefont {Liu}}, \bibinfo
  {author} {\bibfnamefont {Z.-K.}\ \bibnamefont {Guo}}, \bibinfo {author}
  {\bibfnamefont {Y.-L.}\ \bibnamefont {Wu}},\ and\ \bibinfo {author}
  {\bibfnamefont {R.-G.}\ \bibnamefont {Cai}},\ }\bibinfo {title} {{The
  LISA-Taiji network: precision localization of massive black hole binaries}},\
  \Eprint {https://arxiv.org/abs/1909.07104} {arXiv:1909.07104 [gr-qc]}
  \BibitemShut {NoStop}%
\bibitem [{\citenamefont {Ruan}\ {\it et~al.}(2020)\citenamefont {Ruan},
  \citenamefont {Liu}, \citenamefont {Guo}, \citenamefont {Wu},\ and\
  \citenamefont {Cai}}]{Ruan:2020smc}%
  \BibitemOpen
  \bibfield  {author} {\bibinfo {author} {\bibfnamefont {W.-H.}\ \bibnamefont
  {Ruan}}, \bibinfo {author} {\bibfnamefont {C.}~\bibnamefont {Liu}}, \bibinfo
  {author} {\bibfnamefont {Z.-K.}\ \bibnamefont {Guo}}, \bibinfo {author}
  {\bibfnamefont {Y.-L.}\ \bibnamefont {Wu}},\ and\ \bibinfo {author}
  {\bibfnamefont {R.-G.}\ \bibnamefont {Cai}},\ }\bibinfo {title} {{The
  LISA-Taiji network}},\ \href {https://doi.org/10.1038/s41550-019-1008-4}
  {\bibfield  {journal} {\bibinfo  {journal} {Nature Astron.}\ }\textbf
  {\bibinfo {volume} {4}},\ \bibinfo {pages} {108} (\bibinfo {year}
  {2020})}\BibitemShut {NoStop}%
\bibitem [{\citenamefont {Wang}\ {\it et~al.}(2020)\citenamefont {Wang},
  \citenamefont {Ni}, \citenamefont {Han}, \citenamefont {Yang},\ and\
  \citenamefont {Zhong}}]{Wang:2020vkg}%
  \BibitemOpen
  \bibfield  {author} {\bibinfo {author} {\bibfnamefont {G.}~\bibnamefont
  {Wang}}, \bibinfo {author} {\bibfnamefont {W.-T.}\ \bibnamefont {Ni}},
  \bibinfo {author} {\bibfnamefont {W.-B.}\ \bibnamefont {Han}}, \bibinfo
  {author} {\bibfnamefont {S.-C.}\ \bibnamefont {Yang}},\ and\ \bibinfo
  {author} {\bibfnamefont {X.-Y.}\ \bibnamefont {Zhong}},\ }\bibinfo {title}
  {{Numerical simulation of sky localization for LISA-TAIJI joint
  observation}},\ \href {https://doi.org/10.1103/PhysRevD.102.024089}
  {\bibfield  {journal} {\bibinfo  {journal} {Phys. Rev. D}\ }\textbf {\bibinfo
  {volume} {102}},\ \bibinfo {pages} {024089} (\bibinfo {year}
  {2020})}\BibitemShut {NoStop}%
\bibitem [{\citenamefont {Feng}\ {\it et~al.}(2019)\citenamefont {Feng},
  \citenamefont {Wang}, \citenamefont {Hu}, \citenamefont {Hu},\ and\
  \citenamefont {Wang}}]{Feng:2019wgq}%
  \BibitemOpen
  \bibfield  {author} {\bibinfo {author} {\bibfnamefont {W.-F.}\ \bibnamefont
  {Feng}}, \bibinfo {author} {\bibfnamefont {H.-T.}\ \bibnamefont {Wang}},
  \bibinfo {author} {\bibfnamefont {X.-C.}\ \bibnamefont {Hu}}, \bibinfo
  {author} {\bibfnamefont {Y.-M.}\ \bibnamefont {Hu}},\ and\ \bibinfo {author}
  {\bibfnamefont {Y.}~\bibnamefont {Wang}},\ }\bibinfo {title} {{Preliminary
  study on parameter estimation accuracy of supermassive black hole binary
  inspirals for TianQin}},\ \href {https://doi.org/10.1103/PhysRevD.99.123002}
  {\bibfield  {journal} {\bibinfo  {journal} {Phys. Rev. D}\ }\textbf {\bibinfo
  {volume} {99}},\ \bibinfo {pages} {123002} (\bibinfo {year}
  {2019})}\BibitemShut {NoStop}%
\bibitem [{\citenamefont {Huang}\ {\it et~al.}(2020)\citenamefont {Huang},
  \citenamefont {Hu}, \citenamefont {Korol}, \citenamefont {Li}, \citenamefont
  {Liang}, \citenamefont {Lu}, \citenamefont {Wang}, \citenamefont {Yu},\ and\
  \citenamefont {Mei}}]{Huang:2020rjf}%
  \BibitemOpen
  \bibfield  {author} {\bibinfo {author} {\bibfnamefont {S.-J.}\ \bibnamefont
  {Huang}}, \bibinfo {author} {\bibfnamefont {Y.-M.}\ \bibnamefont {Hu}},
  \bibinfo {author} {\bibfnamefont {V.}~\bibnamefont {Korol}}, \bibinfo
  {author} {\bibfnamefont {P.-C.}\ \bibnamefont {Li}}, \bibinfo {author}
  {\bibfnamefont {Z.-C.}\ \bibnamefont {Liang}}, \bibinfo {author}
  {\bibfnamefont {Y.}~\bibnamefont {Lu}}, \bibinfo {author} {\bibfnamefont
  {H.-T.}\ \bibnamefont {Wang}}, \bibinfo {author} {\bibfnamefont
  {S.}~\bibnamefont {Yu}},\ and\ \bibinfo {author} {\bibfnamefont
  {J.}~\bibnamefont {Mei}},\ }\bibinfo {title} {{Science with the TianQin
  Observatory: Preliminary results on Galactic double white dwarf binaries}},\
  \href {https://doi.org/10.1103/PhysRevD.102.063021} {\bibfield  {journal}
  {\bibinfo  {journal} {Phys. Rev. D}\ }\textbf {\bibinfo {volume} {102}},\
  \bibinfo {pages} {063021} (\bibinfo {year} {2020})}\BibitemShut {NoStop}%
\bibitem [{\citenamefont {Zhang}\ {\it et~al.}(2020)\citenamefont {Zhang},
  \citenamefont {Gong}, \citenamefont {Liu}, \citenamefont {Wang},\ and\
  \citenamefont {Zhang}}]{Zhang:2020hyx}%
  \BibitemOpen
  \bibfield  {author} {\bibinfo {author} {\bibfnamefont {C.}~\bibnamefont
  {Zhang}}, \bibinfo {author} {\bibfnamefont {Y.}~\bibnamefont {Gong}},
  \bibinfo {author} {\bibfnamefont {H.}~\bibnamefont {Liu}}, \bibinfo {author}
  {\bibfnamefont {B.}~\bibnamefont {Wang}},\ and\ \bibinfo {author}
  {\bibfnamefont {C.}~\bibnamefont {Zhang}},\ }\bibinfo {title} {{Sky
  localization of space-based gravitational wave detectors}},\ \Eprint
  {https://arxiv.org/abs/2009.03476} {arXiv:2009.03476 [astro-ph.IM]}
  \BibitemShut {NoStop}%
\bibitem [{\citenamefont {Vallisneri}(2008)}]{Vallisneri:2007ev}%
  \BibitemOpen
  \bibfield  {author} {\bibinfo {author} {\bibfnamefont {M.}~\bibnamefont
  {Vallisneri}},\ }\bibinfo {title} {{Use and abuse of the Fisher information
  matrix in the assessment of gravitational-wave parameter-estimation
  prospects}},\ \href {https://doi.org/10.1103/PhysRevD.77.042001} {\bibfield
  {journal} {\bibinfo  {journal} {Phys. Rev. D}\ }\textbf {\bibinfo {volume}
  {77}},\ \bibinfo {pages} {042001} (\bibinfo {year} {2008})}\BibitemShut
  {NoStop}%
\bibitem [{\citenamefont {Wen}\ and\ \citenamefont {Chen}(2010)}]{Wen:2010cr}%
  \BibitemOpen
  \bibfield  {author} {\bibinfo {author} {\bibfnamefont {L.}~\bibnamefont
  {Wen}}\ and\ \bibinfo {author} {\bibfnamefont {Y.}~\bibnamefont {Chen}},\
  }\bibinfo {title} {{Geometrical Expression for the Angular Resolution of a
  Network of Gravitational-Wave Detectors}},\ \href
  {https://doi.org/10.1103/PhysRevD.81.082001} {\bibfield  {journal} {\bibinfo
  {journal} {Phys. Rev. D}\ }\textbf {\bibinfo {volume} {81}},\ \bibinfo
  {pages} {082001} (\bibinfo {year} {2010})}\BibitemShut {NoStop}%
\bibitem [{\citenamefont {Abbott}\ {\it et~al.}(2018)\citenamefont {Abbott}
  {\it et~al.}}]{Aasi:2013wya}%
  \BibitemOpen
  \bibfield  {author} {\bibinfo {author} {\bibfnamefont {B.~P.}\ \bibnamefont
  {Abbott}} {\it et~al.} (\bibinfo {collaboration} {KAGRA, LIGO Scientific,
  VIRGO}),\ }\bibinfo {title} {{Prospects for Observing and Localizing
  Gravitational-Wave Transients with Advanced LIGO, Advanced Virgo and
  KAGRA}},\ \href {https://doi.org/10.1007/s41114-018-0012-9} {\bibfield
  {journal} {\bibinfo  {journal} {Living Rev. Rel.}\ }\textbf {\bibinfo
  {volume} {21}},\ \bibinfo {pages} {3} (\bibinfo {year} {2018})}\BibitemShut
  {NoStop}%
\bibitem [{\citenamefont {Grover}\ {\it et~al.}(2014)\citenamefont {Grover},
  \citenamefont {Fairhurst}, \citenamefont {Farr}, \citenamefont {Mandel},
  \citenamefont {Rodriguez}, \citenamefont {Sidery},\ and\ \citenamefont
  {Vecchio}}]{Grover:2013sha}%
  \BibitemOpen
  \bibfield  {author} {\bibinfo {author} {\bibfnamefont {K.}~\bibnamefont
  {Grover}}, \bibinfo {author} {\bibfnamefont {S.}~\bibnamefont {Fairhurst}},
  \bibinfo {author} {\bibfnamefont {B.~F.}\ \bibnamefont {Farr}}, \bibinfo
  {author} {\bibfnamefont {I.}~\bibnamefont {Mandel}}, \bibinfo {author}
  {\bibfnamefont {C.}~\bibnamefont {Rodriguez}}, \bibinfo {author}
  {\bibfnamefont {T.}~\bibnamefont {Sidery}},\ and\ \bibinfo {author}
  {\bibfnamefont {A.}~\bibnamefont {Vecchio}},\ }\bibinfo {title} {{Comparison
  of Gravitational Wave Detector Network Sky Localization Approximations}},\
  \href {https://doi.org/10.1103/PhysRevD.89.042004} {\bibfield  {journal}
  {\bibinfo  {journal} {Phys. Rev. D}\ }\textbf {\bibinfo {volume} {89}},\
  \bibinfo {pages} {042004} (\bibinfo {year} {2014})}\BibitemShut {NoStop}%
\bibitem [{\citenamefont {Berry}\ {\it et~al.}(2015)\citenamefont {Berry} {\it
  et~al.}}]{Berry:2014jja}%
  \BibitemOpen
  \bibfield  {author} {\bibinfo {author} {\bibfnamefont {C.~P.~L.}\
  \bibnamefont {Berry}} {\it et~al.},\ }\bibinfo {title} {{Parameter estimation
  for binary neutron-star coalescences with realistic noise during the Advanced
  LIGO era}},\ \href {https://doi.org/10.1088/0004-637X/804/2/114} {\bibfield
  {journal} {\bibinfo  {journal} {Astrophys. J.}\ }\textbf {\bibinfo {volume}
  {804}},\ \bibinfo {pages} {114} (\bibinfo {year} {2015})}\BibitemShut
  {NoStop}%
\bibitem [{\citenamefont {Singer}\ and\ \citenamefont
  {Price}(2016)}]{Singer:2015ema}%
  \BibitemOpen
  \bibfield  {author} {\bibinfo {author} {\bibfnamefont {L.~P.}\ \bibnamefont
  {Singer}}\ and\ \bibinfo {author} {\bibfnamefont {L.~R.}\ \bibnamefont
  {Price}},\ }\bibinfo {title} {{Rapid Bayesian position reconstruction for
  gravitational-wave transients}},\ \href
  {https://doi.org/10.1103/PhysRevD.93.024013} {\bibfield  {journal} {\bibinfo
  {journal} {Phys. Rev. D}\ }\textbf {\bibinfo {volume} {93}},\ \bibinfo
  {pages} {024013} (\bibinfo {year} {2016})}\BibitemShut {NoStop}%
\bibitem [{\citenamefont {B\'ecsy}\ {\it et~al.}(2017)\citenamefont {B\'ecsy},
  \citenamefont {Raffai}, \citenamefont {Cornish}, \citenamefont {Essick},
  \citenamefont {Kanner}, \citenamefont {Katsavounidis}, \citenamefont
  {Littenberg}, \citenamefont {Millhouse},\ and\ \citenamefont
  {Vitale}}]{Becsy:2016ofp}%
  \BibitemOpen
  \bibfield  {author} {\bibinfo {author} {\bibfnamefont {B.}~\bibnamefont
  {B\'ecsy}}, \bibinfo {author} {\bibfnamefont {P.}~\bibnamefont {Raffai}},
  \bibinfo {author} {\bibfnamefont {N.~J.}\ \bibnamefont {Cornish}}, \bibinfo
  {author} {\bibfnamefont {R.}~\bibnamefont {Essick}}, \bibinfo {author}
  {\bibfnamefont {J.}~\bibnamefont {Kanner}}, \bibinfo {author} {\bibfnamefont
  {E.}~\bibnamefont {Katsavounidis}}, \bibinfo {author} {\bibfnamefont {T.~B.}\
  \bibnamefont {Littenberg}}, \bibinfo {author} {\bibfnamefont
  {M.}~\bibnamefont {Millhouse}},\ and\ \bibinfo {author} {\bibfnamefont
  {S.}~\bibnamefont {Vitale}},\ }\bibinfo {title} {{Parameter estimation for
  gravitational-wave bursts with the BayesWave pipeline}},\ \href
  {https://doi.org/10.3847/1538-4357/aa63ef} {\bibfield  {journal} {\bibinfo
  {journal} {Astrophys. J.}\ }\textbf {\bibinfo {volume} {839}},\ \bibinfo
  {pages} {15} (\bibinfo {year} {2017})}\BibitemShut {NoStop}%
\bibitem [{\citenamefont {Zhao}\ and\ \citenamefont
  {Wen}(2018)}]{Zhao:2017cbb}%
  \BibitemOpen
  \bibfield  {author} {\bibinfo {author} {\bibfnamefont {W.}~\bibnamefont
  {Zhao}}\ and\ \bibinfo {author} {\bibfnamefont {L.}~\bibnamefont {Wen}},\
  }\bibinfo {title} {{Localization accuracy of compact binary coalescences
  detected by the third-generation gravitational-wave detectors and implication
  for cosmology}},\ \href {https://doi.org/10.1103/PhysRevD.97.064031}
  {\bibfield  {journal} {\bibinfo  {journal} {Phys. Rev. D}\ }\textbf {\bibinfo
  {volume} {97}},\ \bibinfo {pages} {064031} (\bibinfo {year}
  {2018})}\BibitemShut {NoStop}%
\bibitem [{\citenamefont {Mills}\ {\it et~al.}(2018)\citenamefont {Mills},
  \citenamefont {Tiwari},\ and\ \citenamefont {Fairhurst}}]{Mills:2017urp}%
  \BibitemOpen
  \bibfield  {author} {\bibinfo {author} {\bibfnamefont {C.}~\bibnamefont
  {Mills}}, \bibinfo {author} {\bibfnamefont {V.}~\bibnamefont {Tiwari}},\ and\
  \bibinfo {author} {\bibfnamefont {S.}~\bibnamefont {Fairhurst}},\ }\bibinfo
  {title} {{Localization of binary neutron star mergers with second and third
  generation gravitational-wave detectors}},\ \href
  {https://doi.org/10.1103/PhysRevD.97.104064} {\bibfield  {journal} {\bibinfo
  {journal} {Phys. Rev. D}\ }\textbf {\bibinfo {volume} {97}},\ \bibinfo
  {pages} {104064} (\bibinfo {year} {2018})}\BibitemShut {NoStop}%
\bibitem [{\citenamefont {Fairhurst}(2018)}]{Fairhurst:2017mvj}%
  \BibitemOpen
  \bibfield  {author} {\bibinfo {author} {\bibfnamefont {S.}~\bibnamefont
  {Fairhurst}},\ }\bibinfo {title} {{Localization of transient gravitational
  wave sources: beyond triangulation}},\ \href
  {https://doi.org/10.1088/1361-6382/aab675} {\bibfield  {journal} {\bibinfo
  {journal} {Class. Quant. Grav.}\ }\textbf {\bibinfo {volume} {35}},\ \bibinfo
  {pages} {105002} (\bibinfo {year} {2018})}\BibitemShut {NoStop}%
\bibitem [{\citenamefont {Fujii}\ {\it et~al.}(2019)\citenamefont {Fujii},
  \citenamefont {Adams}, \citenamefont {Marion},\ and\ \citenamefont
  {Flaminio}}]{Fujii:2019hdi}%
  \BibitemOpen
  \bibfield  {author} {\bibinfo {author} {\bibfnamefont {Y.}~\bibnamefont
  {Fujii}}, \bibinfo {author} {\bibfnamefont {T.}~\bibnamefont {Adams}},
  \bibinfo {author} {\bibfnamefont {F.}~\bibnamefont {Marion}},\ and\ \bibinfo
  {author} {\bibfnamefont {R.}~\bibnamefont {Flaminio}},\ }\bibinfo {title}
  {{Fast localization of coalescing binaries with a heterogeneous network of
  advanced gravitational wave detectors}},\ \href
  {https://doi.org/10.1016/j.astropartphys.2019.04.008} {\bibfield  {journal}
  {\bibinfo  {journal} {Astropart. Phys.}\ }\textbf {\bibinfo {volume} {113}},\
  \bibinfo {pages} {1} (\bibinfo {year} {2019})}\BibitemShut {NoStop}%
\bibitem [{\citenamefont {Rubbo}\ {\it et~al.}(2004)\citenamefont {Rubbo},
  \citenamefont {Cornish},\ and\ \citenamefont {Poujade}}]{Rubbo:2003ap}%
  \BibitemOpen
  \bibfield  {author} {\bibinfo {author} {\bibfnamefont {L.~J.}\ \bibnamefont
  {Rubbo}}, \bibinfo {author} {\bibfnamefont {N.~J.}\ \bibnamefont {Cornish}},\
  and\ \bibinfo {author} {\bibfnamefont {O.}~\bibnamefont {Poujade}},\
  }\bibinfo {title} {{Forward modeling of space borne gravitational wave
  detectors}},\ \href {https://doi.org/10.1103/PhysRevD.69.082003} {\bibfield
  {journal} {\bibinfo  {journal} {Phys. Rev. D}\ }\textbf {\bibinfo {volume}
  {69}},\ \bibinfo {pages} {082003} (\bibinfo {year} {2004})}\BibitemShut
  {NoStop}%
\bibitem [{\citenamefont {Cornish}\ and\ \citenamefont
  {Larson}(2001)}]{Cornish:2001qi}%
  \BibitemOpen
  \bibfield  {author} {\bibinfo {author} {\bibfnamefont {N.~J.}\ \bibnamefont
  {Cornish}}\ and\ \bibinfo {author} {\bibfnamefont {S.~L.}\ \bibnamefont
  {Larson}},\ }\bibinfo {title} {{Space missions to detect the cosmic
  gravitational wave background}},\ \href
  {https://doi.org/10.1088/0264-9381/18/17/308} {\bibfield  {journal} {\bibinfo
   {journal} {Class. Quant. Grav.}\ }\textbf {\bibinfo {volume} {18}},\
  \bibinfo {pages} {3473} (\bibinfo {year} {2001})}\BibitemShut {NoStop}%
\bibitem [{\citenamefont {Estabrook}\ and\ \citenamefont
  {Wahlquist}(1975)}]{Estabrook:1975}%
  \BibitemOpen
  \bibfield  {author} {\bibinfo {author} {\bibfnamefont {F.~B.}\ \bibnamefont
  {Estabrook}}\ and\ \bibinfo {author} {\bibfnamefont {H.~D.}\ \bibnamefont
  {Wahlquist}},\ }\bibinfo {title} {{Response of Doppler spacecraft tracking to
  gravitational radiation}},\ \href {https://doi.org/10.1007/BF00762449}
  {\bibfield  {journal} {\bibinfo  {journal} {Gen. Relat. Gravit.}\ }\textbf
  {\bibinfo {volume} {6}},\ \bibinfo {pages} {439} (\bibinfo {year}
  {1975})}\BibitemShut {NoStop}%
\bibitem [{\citenamefont {Hu}\ {\it et~al.}(2018)\citenamefont {Hu},
  \citenamefont {Li}, \citenamefont {Wang}, \citenamefont {Feng}, \citenamefont
  {Zhou}, \citenamefont {Hu}, \citenamefont {Hu}, \citenamefont {Mei},\ and\
  \citenamefont {Shao}}]{Hu:2018yqb}%
  \BibitemOpen
  \bibfield  {author} {\bibinfo {author} {\bibfnamefont {X.-C.}\ \bibnamefont
  {Hu}}, \bibinfo {author} {\bibfnamefont {X.-H.}\ \bibnamefont {Li}}, \bibinfo
  {author} {\bibfnamefont {Y.}~\bibnamefont {Wang}}, \bibinfo {author}
  {\bibfnamefont {W.-F.}\ \bibnamefont {Feng}}, \bibinfo {author}
  {\bibfnamefont {M.-Y.}\ \bibnamefont {Zhou}}, \bibinfo {author}
  {\bibfnamefont {Y.-M.}\ \bibnamefont {Hu}}, \bibinfo {author} {\bibfnamefont
  {S.-C.}\ \bibnamefont {Hu}}, \bibinfo {author} {\bibfnamefont {J.-W.}\
  \bibnamefont {Mei}},\ and\ \bibinfo {author} {\bibfnamefont {C.-G.}\
  \bibnamefont {Shao}},\ }\bibinfo {title} {{Fundamentals of the orbit and
  response for TianQin}},\ \href {https://doi.org/10.1088/1361-6382/aab52f}
  {\bibfield  {journal} {\bibinfo  {journal} {Class. Quant. Grav.}\ }\textbf
  {\bibinfo {volume} {35}},\ \bibinfo {pages} {095008} (\bibinfo {year}
  {2018})}\BibitemShut {NoStop}%
\end{thebibliography}

%

\end{document}